\documentclass[emulateapj,twocolumn]{aastex63}
\usepackage{graphicx}
\usepackage{float}
\usepackage{amsmath}
\usepackage{subfigure}
\usepackage{natbib}
\usepackage{array}
\usepackage{gensymb}
\newcommand{\head}[2]{\multicolumn{1}{>{\centering\arraybackslash}p{#1}}{\textbf{#2}}}

\usepackage{booktabs}
 
\begin{document}

\title{ALMA evidence for ram pressure compression and stripping of molecular gas in the Virgo cluster galaxy NGC 4402}
\author{Cramer, W. J.}
\affil{Department of Astronomy, Yale University, New Haven, CT 06511, USA}
\author{Kenney, J.D.P.}
\affil{Department of Astronomy, Yale University, New Haven, CT 06511, USA}
\author{Cortes, J. R.}
\affil{National Radio Astronomy Observatory Avenida Nueva Costanera 4091, Vitacura, Santiago, Chile \footnote{The National Radio Astronomy Observatory is a facility of the National Science Foundation operated under cooperative agreement by Associated Universities, Inc.}}
\affil{Joint ALMA Observatory, Alonso de C{\'o}rdova 3107, Vitacura, Santiago, Chile}
\author{Cortes, P. C.}
\affil{National Radio Astronomy Observatory Avenida Nueva Costanera 4091, Vitacura, Santiago, Chile \footnote{The National Radio Astronomy Observatory is a facility of the National Science Foundation operated under cooperative agreement by Associated Universities, Inc.}}
\affil{Joint ALMA Observatory, Alonso de C{\'o}rdova 3107, Vitacura, Santiago, Chile}
\author{Vlahakis, C.}
\affil{National Radio Astronomy Observatory, 520 Edgemont Road, Charlottesville, VA 22903-2475, USA}
\author{J{\'a}chym, P.}
\affil{Astronomical Institute, Czech Academy of Sciences, Prague, Czech Republic}
\author{Pompei, E.}
\affil{ESO-Chile, Alonso de Cordova 3107, Vitacura, Santiago, Chile}
\author{Rubio, M.}
\affil{Departamento de Astronom{\'i}a, Universidad de Chile, Casilla 36-D, Santiago, Chile}

\begin{abstract}

High resolution (1'' $\times$ 2'') ALMA CO(2-1) observations of the ram pressure stripped galaxy NGC 4402 in the Virgo cluster show some of the clearest evidence yet for the impacts of ram pressure on the molecular ISM of a galaxy. The eastern side of the galaxy at $r \sim 4.5$ kpc, upon which ram pressure is incident, has a large (width $\sim$1 kpc, height $\sim$1 kpc above the disk midplane) extraplanar plume of molecular gas and dust. Molecular gas in the plume region shows distinct non-circular motions in the direction of the ram pressure; the kinematic offset of up to 60 km s$^{-1}$ is consistent with acceleration by ram pressure. We also detect a small amount of gas in clouds below the plume that are spatially and kinematically distinct from the surrounding medium, and appear to be decoupled from the stripped ISM. We propose that diffuse molecular gas is directly stripped but GMC density gas is not directly stripped, and so decouples from lower density stripped gas. However, GMCs become effectively stripped on short timescales. We also find morphological and kinematic signatures of ram pressure compression of molecular gas in a region of intense star formation on the leading side at $r \sim 3.5$ kpc. We propose that the compressed and stripped zones represent different evolutionary stages of the ram pressure interaction, and that feedback from star formation in the compressed zone facilitates the effective stripping of GMCs by making the gas cycle rapidly to a lower density diffuse state.

\end{abstract}

\section{Introduction}

Ram pressure stripping (RPS) is a major source of star formation quenching in galaxies in clusters, as has been identified by the truncated gas disks with one-sided gas tails and the outside-in radial quenching signature it leaves as a result of selective outside-in removal of gas. \citep{Pappalardo+10, Abramson+11, Merluzzi+16, Fossati+18, Cramer+19}. Ram pressure can also trigger short-lived starbursts before the gas is removed \citep{Crowl+06, Vollmer+12, Vulcani+18}, and in some cases these starbursts form more massive star clusters \citep{Lee+16}. In order to understand the general importance of RPS in galaxy evolution, including in environments such as groups \citep{Smethurst+17, Vulcani+18}, galaxy pairs \citep{Moon+19}, and in the earlier universe \citep{Mayer+06, Boselli+19}, we need to understand how easily gas is stripped, and the physical conditions that produce either triggered star formation or stripping. The impact of RPS on the evolution of galaxies depends on the behavior of the densest gas, i.e. the molecular gas, during stripping episodes, as it is the medium in which stars form, and in some circumstances is the dominant interstellar medium (ISM) component by mass.

According to the simple Gunn \& Gott formula \citep{Gunn+72}, which compares the ram pressure force to the gravitational restoring force, stripping susceptibility is linearly proportional to the gas surface density $\Sigma_{gas}$, meaning that molecular gas, which is the densest gas, is least susceptible to stripping. The relevant physics are undoubtedly more complex than is implied by the Gunn \& Gott relation, but additional factors would not change the fact that denser gas is harder to strip.

If stripping susceptibility is well approximated by the Gunn \& Gott relation, and the standard estimates of the intercluster medium (ICM) gas density and relative velocity are close to correct, giant molecular clouds like those in the solar neighborhood of the Milky Way, with gas surface densities of $\sim$100 M$_{\odot}$ pc$^2$ \citep{Blitz+06}, should resist stripping in intermediate mass clusters like Virgo (see \citet{Crowl+05, Lee+17} for examples of these estimates). Simulations by \citet{Tonnesen+09, Tonnesen+10} of the ISM in galaxies under ram pressure comparable to that in the Virgo cluster also find no evidence for direct stripping of gas with the densities typical of giant molecular clouds (GMCs). While significant amounts of molecular gas have been found in the extraplanar gas tails of ram pressure stripped galaxies \citep{Vollmer+08, Sun+10, Jachym+14, Jachym+17}, the gas has been generally assumed to cool and form in situ. It is possible however, that in more massive clusters such as Coma where the average strength of ram pressure is estimated to be approximately $10-100$ times higher than Virgo, some direct stripping of GMCs may occur, although it has never been directly observed.

Early studies found no CO deficiency in HI deficient Virgo cluster spirals \citep{Kenney+89}, which was attributed to GMCs being too dense to strip in Virgo, although the reality is more complex. In \citet{Boselli+14}, the authors found a correlation between HI deficiency and modest CO deficiency in a survey of Virgo cluster galaxies. The large HI and modest molecular gas deficiencies of cluster galaxies can be understood together if nearly all the gas beyond some radius is stripped \citep{Cortese+16}. Gas from the inner galaxy is harder to strip, and molecular gas is more centrally concentrated than HI. This corresponds well with the observation of well-defined gas truncation radii in most stripped Virgo spirals, with almost no dust extinction, HI, HII regions, or star formation beyond a fairly well-defined gas truncation radius \citep{Cayatte+90, Chung+09, Koopmann+04, Cramer+19}. Apparently when the ISM gets stripped, virtually all the gas, including the molecular gas, seems to disappear fairly quickly, and molecular gas is thus “effectively stripped” \citep{Kenney+04, Vollmer+12d, Boselli+14}, despite the fact that GMCs should be too dense to be directly stripped in Virgo. Some gas may still remain in the stripped zone, as found by \citet{Vollmer+05, Vollmer+12d} from CO observations of NGC 4438, and as evidenced by decoupled clouds of dust observed just beyond the main gas truncation radius in Hubble Space Telescope (HST) images of two actively stripped Virgo spirals \citep{Abramson+14}, and these are proposed to be clouds which are too dense to directly strip. However, only a small fraction of the pre-stripped molecular gas mass is found in these clouds, so most of the molecular gas is effectively stripped.

The physical mechanism by which RPS results in the removal of most of the molecular gas remains unclear. Of critical importance is the typical lifetime of a GMC, and the relative amounts of molecular gas in the diffuse and GMC states. \citet{Roman+16} found that $\sim$25\% of the molecular gas in the Milky Way is diffuse, and this fraction increases with radial distance from the galaxy center. In the spiral galaxy M51, \citet{Pety+13} found that $\sim$ 50\% of the emission from CO is recovered only on spatial scales larger than $\sim$ 1.3 kpc, which they claim is consistent with this emission being associated with an extended, diffuse disk of molecular gas. Given the large fraction of diffuse molecular gas found by \citet{Roman+16}, it is important to understand whether, and to what degree, diffuse molecular gas is stripped in galaxies, and how this affects the overall evolution of the multiphase ISM under stripping. 

To date, no convincing evidence has been found showing the direct transport of diffuse or GMC density molecular gas by ram pressure, particularly with a clear velocity signature of the gas transport. While \citep{Moretti+20} found significant displacement between the molecular gas and stellar disk in the ram pressure stripped galaxy JW100, the orientation of the galaxy and resolution of the observations made a kinematic analysis showing clear molecular gas transport via ram pressure impossible. Only the resolution of the Atacama Large Millimeter Array (ALMA) has made it possible to study the detailed gas distribution with sufficient spatial and velocity resolution to observe the non-circular motions in the molecular gas caused by RPS, as the gas is accelerated away from the disk.

\subsection{NGC 4402}

\begin{figure}
	\plotone{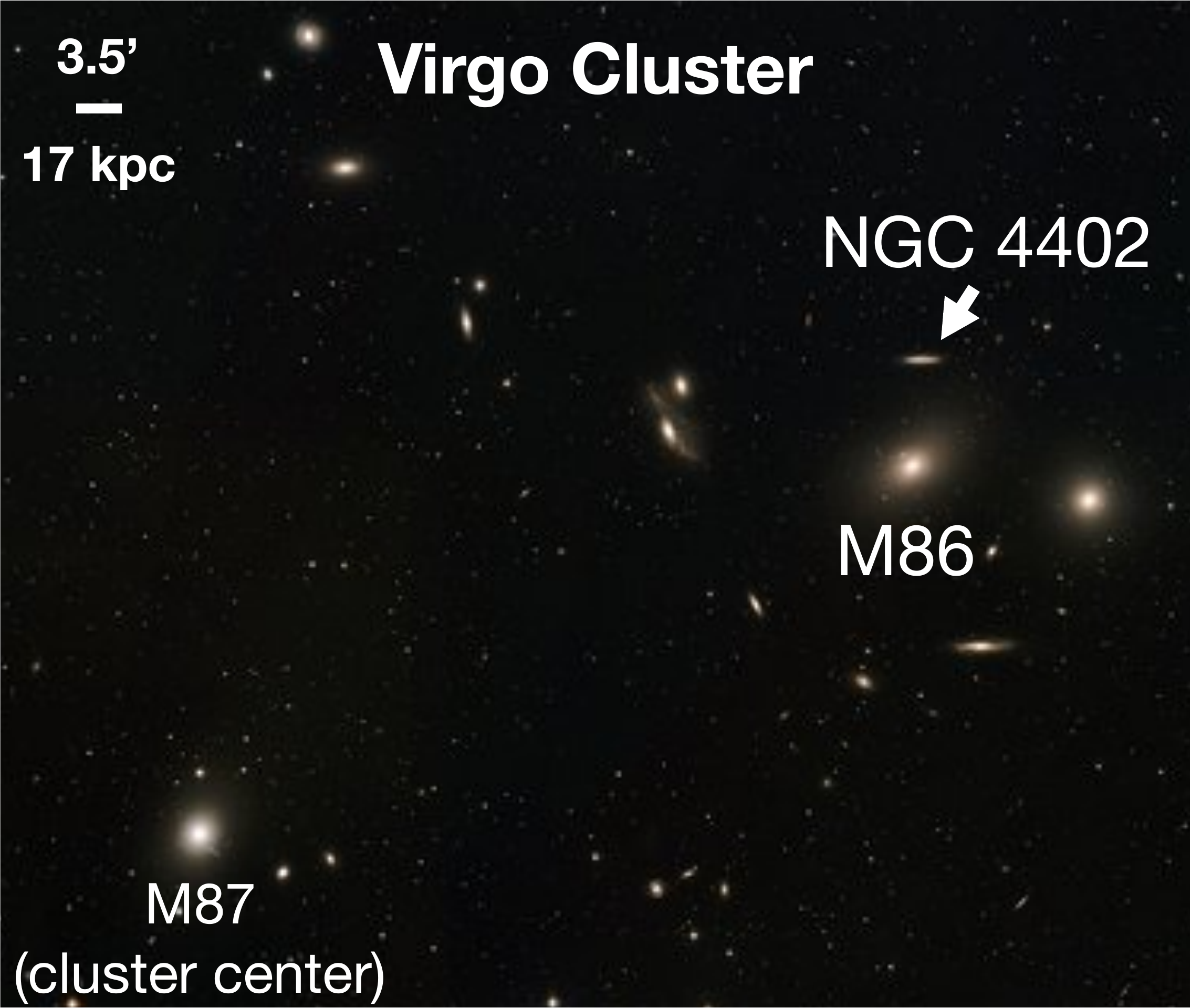}
	\caption{A 4.2 degrees x 4.2 degrees image of the region around NGC 4402 constructed from images in the Digitized Sky Survey (from NASA/ESA (\url{https://spacetelescope.org/images/heic0911f/}). NGC 4402 and the nearby M86 subcluster are indicated, as well as M87, located at the cluster center.}
	\label{fig:Virgo_cluster}
\end{figure}

\begin{figure*}
	\plotone{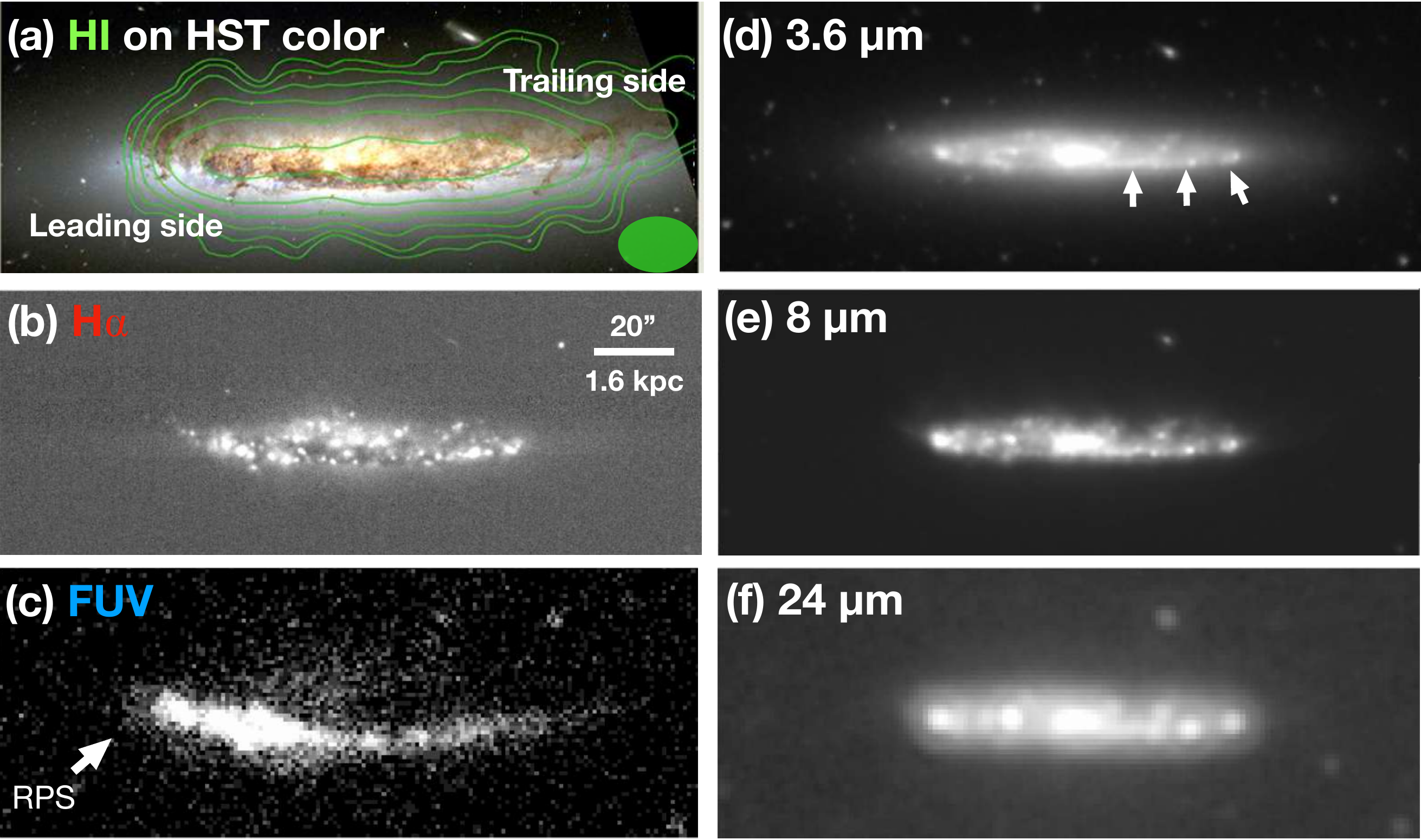}
	\caption{\textbf{(a):} HI (with contours varying from 0.4 to 3.2 mJy by factors of two) on an HST B, V, I color image, from \citet{Abramson+16}, with the leading side experiencing the strongest RP, and the trailing side, which is more shielded, both labeled. The beam size (20'' $\times$ 17'') is indicated by the green oval in the lower right. \textbf{(b):} A continuum subtracted H$\alpha$ image of NGC 4402 \citep{Kenney+08}. \textbf{(c):} A GALEX FUV image of NGC 4402 \citep{Paz+07}. The estimated direction of RPS based on the projected motion of the galaxy is also shown. \textbf{(d), (e), (f):} From top to bottom, Spitzer  3.6, 8, and 24 $\mu$m images of NGC 4402 \citep{Kenney+09}. The 3.6 $\mu$m traces the stellar component, and we also indicate what looks like a strong spiral arm feature on the west side of the galaxy. 8 $\mu$m flux is more sensitive to emission from PAH emission, and 24 $\mu$m flux is an indicator of star formation.}
	\label{fig:SF_tracers}
\end{figure*}

NGC 4402 is one of the nearest and clearest examples of a spiral galaxy experiencing ram pressure stripping. It is close enough that one can study with high spatial and kinematic resolution the effects of RPS on gas in the disk. NGC 4402 is located in the Virgo cluster at a distance of $\sim$17 Mpc \citep{Mei+07}, and $\sim$0.4 Mpc in projected distance from the cluster center, as indicated in Figure \ref{fig:Virgo_cluster}. It may be a member of the merging subcluster located around the giant elliptical M86 \citep{Bohringer+94}. Images of the galaxy at different wavelengths shown in Figure \ref{fig:SF_tracers} reveal a highly inclined ($i=80$ deg), 0.3L* Sc galaxy. Infrared images at 3.6, 8, and 24$\mu$m (Figure \ref{fig:SF_tracers}d, e, f) show a spiral arm on the western side of the galaxy, but more irregular structure on the east. Tracers of young stars, e.g. 24$\mu$m, H$\alpha$, FUV, (Figure \ref{fig:SF_tracers}b, c, f) show emission truncated at a radius of 75” (6 kpc), indicating that star formation has been quenched beyond here.

\citet{Crowl+05} established NGC 4402 as being actively ram pressure stripped, via the detection of a prominent radio continuum tail. They also identified several large filaments of dust in the galaxy that protrude into the otherwise stripped zone, and use the morphology of these features to estimate the projected wind direction. Furthermore, \citet{Crowl+05} also found the stellar disk of NGC 4402 to be undisturbed, ruling out any strong influence from tidal interactions on the structure of the galaxy. Some of the effects of the RPS on the ISM of the galaxy were characterized through analysis of the dust extinction features with HST \citep{Abramson+14, Abramson+16}. The authors found dense clouds of dust, comprising $\sim 1\%$ of the pre-stripped ISM mass, that appeared to be the only surviving ISM in the mostly stripped, HI deficient southern zone of the galaxy. 

The galaxy was also observed in HI with the VLA by \citet{Crowl+05}, as part of the survey of Virgo galaxies in Atomic gas (VIVA) presented in \citet{Chung+09}. The galaxy is moderately HI deficient, with a deficiency parameter of 0.61, which corresponds to about 25\% of the typical HI content for a galaxy in a low-density environment \citet{Giovanelli+83}. We reproduce the VIVA HI map in Figure \ref{fig:SF_tracers}a to show that the HI appears to be compressed and truncated on the leading side of the galaxy at about the same radius as the star formation is truncated \citep{Crowl+05, Chung+09}. NGC 4402 also has a modest HI tail that can be seen on the trailing side. HST images of the galaxy shown in Figure \ref{fig:SF_tracers}a shows a dust plume rising above the disk on the leading (eastern) side. While the HI appears approximately coextensive with the dust, due to the low resolution (15'' $\sim$ 1.2 kpc) of the VLA HI observations, it is not possible to tell how much is in the disk and how much is extraplanar, or to begin to resolve any of the disk ISM features affected by ram pressure. 

Star formation tracers (H$\alpha$, 24$\mu$m, and particularly FUV) show a local peak at the leading edge of the disk in the east, as seen in Figure \ref{fig:SF_tracers}. It is possible this is star formation which has been triggered by ram pressure (see further discussion in section \ref{Evolution}).

NGC 4402 was previously observed in CO (1-0) by \citet{Kenney+89}, in which the authors found a roughly normal CO content, but an asymmetric gas distribution, with stronger emission on the eastern side. The galaxy was recently observed in CO (2-1) by \citet{Lee+17} with the Submillimeter Array (SMA) at a resolution of 7'' $\times$ 4''. The authors compared their observations with HI maps and found that, like the HI, CO and H$\alpha$ appeared to be compressed on the southern side of the disk and more extended to the north, consistent with ram pressure from the southeast.

Our ALMA observations reveal a greater spatial extent of CO emission, and much more detail related to the ram pressure interaction than could be seen with the SMA data of \citet{Lee+17} as we have greater resolution and sensitivity.

\subsection{Outline}

In this paper we present a detailed study of the molecular gas distribution and kinematics of
NGC 4402 at 1'' $\times$ 2'' (80 $\times$ 160 pc) resolution, focusing on the eastern (leading) side since that is the region most affected by ram pressure. In section 2 we present technical details of our observations. In section 3, we present the data and provide a general description of the overall CO distribution and kinematics. In section 4 we present our velocity model of the galaxy, and with this we identify non-circular motions in the data that are presumably due to ram pressure. In section 5, we analyze the kinematics, masses, densities, spatial extents, and other key properties of interesting ISM features sculpted by ram pressure at the leading side. In section 6, we discuss the implications of our findings, and the predicted evolution of the ISM as a result of ram pressure, via compression, star formation, and stripping. Finally, in section 7 we present a summary of our results.

We plan to present ALMA CO(3-2) observations and discuss line ratios as well as the western side of the galaxy in Paper II.

\section{Observations}

We observed NGC 4402 with ALMA in 2015-16 during Cycle 3 \& 4, (project codes 2015.1.01056.S, 2016.1.00912.S, PI: Kenney). We observed in Band 6, centered around the CO(2-1) line at 230.36 GHz. We also observed the CS(5-4) line, but found no emission.

The galaxy was mapped in mosaicking mode with 67 pointings using the 12m array, and 25 pointings using the 7m ALMA Compact Array (ACA) to recover emission on larger angular scales. Their respective primary beams are 25'' for the 12m antenna and 43'' for the 7m antenna. The resulting mosaic has a total angular size of 180'' $\times$ 45'' (1''$ \approx 80$ pc in Virgo). The velocity resolution of the data is 1.27 km s$^{-1}$ and the total time on source is 87.5 minutes with the 12m array and 343 minutes with the ACA. We observed in Band 6 with four spectral windows, each with a width of 1.875GHz. We placed one spectral window on the CO(2-1) line centered at $\sim$230 GHz, another on the CS(5-4) line centered at $\sim$245 GHz, and two final spectral windows used for continuum only centered at 228.5 and 246.5 GHz.

Flux calibration is based on observations of Callisto and Ganymede (ACA) and J1229+0203 (12m), bandpass calibration uses J1229+0203 (12m \& ACA), and the phase calibrators are J1215+1654 (12m) and J1229+0203 (ACA).

The data have been calibrated using the CASA pipeline, and imaged using the CASA software package (version 5.4.0). No significant continuum emission was found in the spectral windows around the CO(2-1) line. We chose to weight the data with a Briggs robustness of 2.0 (closest to natural weighting), with a resulting beam size of  2.3'' $\times$ 1.1''. The CO(2-1) data has an elongated beam due to non-uniform East-West coverage during the observations.

\begin{figure*}
	\plotone{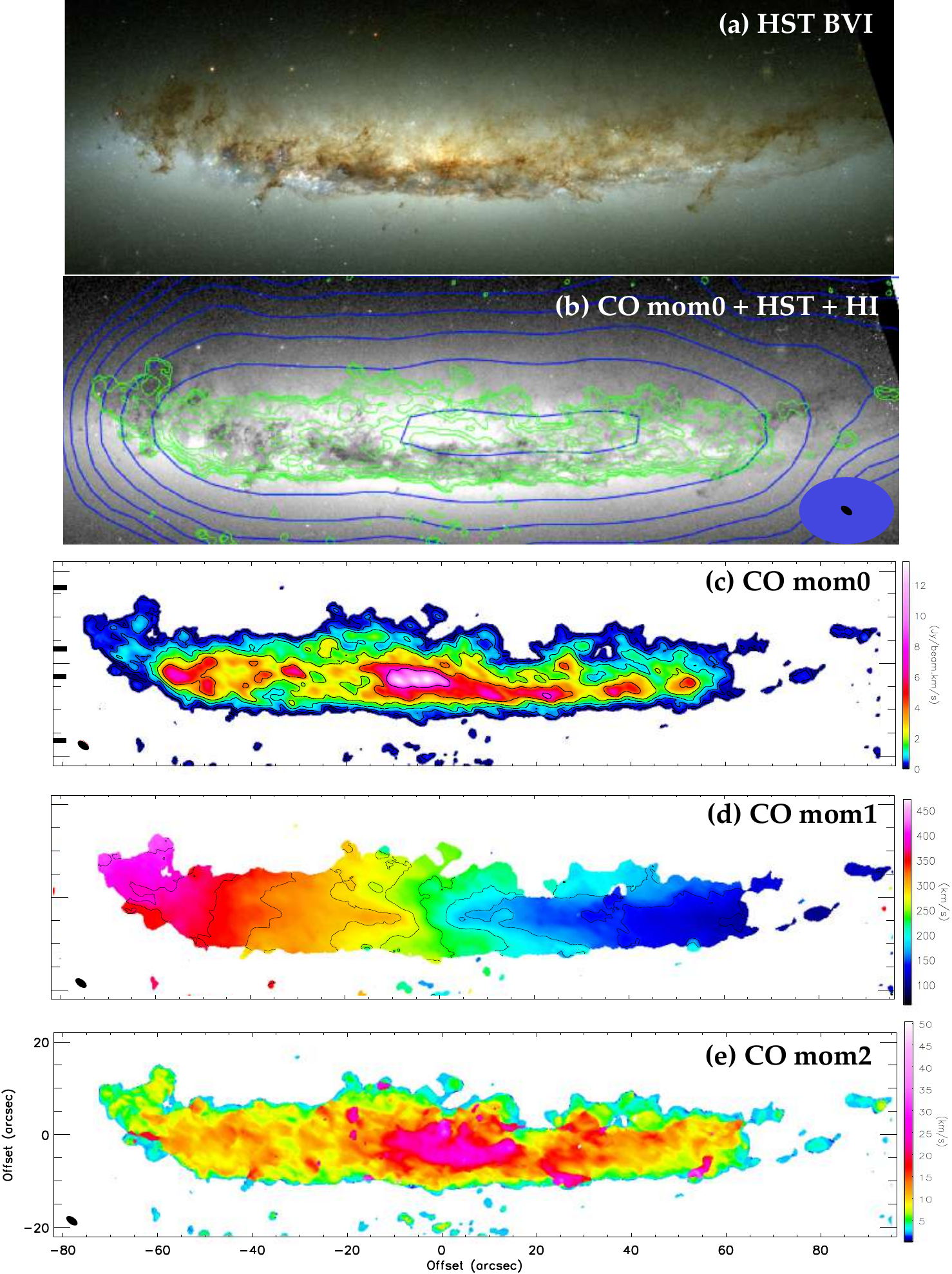}
	\caption{\textbf{(a)}: An HST color image of NGC 4402 made with three filters, F435W in blue, F505W in green, and F814W in red (From a HST press release (\url{https://www.spacetelescope.org/images/heic0911c/}), with data from \citet{Abramson+14}). \textbf{(b)}: CO intensity in green on F606W  with the contour levels ranging from $0.12 - 7.6$ Jy beam$^{-1}$ km s$^{-1}$ or $\sim$$4 - 265$ M$_{\odot}$ pc$^{-2}$, and HI in blue contours with contour levels varying from 12 to 400 mJy beam$^{-1}$ km s$^{-1}$ by factors of two. The ALMA beam is shown as a black filled ellipse on the lower right, and the VLA beam is in blue. \textbf{(c)}: A moment 0 map of our ALMA CO (2-1) data. We note that there is a region at a distance of 15-30'' to the south of the galaxy with extra noise that was not fully removed by cleaning, but we do not believe it corresponds to real emission. We mark the two widths of PVDs referenced later in this work with black marks on the y axis. \textbf{(d)}: A moment 1 velocity map of the CO (2-1). \textbf{(e)}: A moment 2 (velocity dispersion) map of the CO (2-1).
	}
	\label{fig:CO21_mirror}
\end{figure*}

The data were cleaned using the CASA \textit{tclean} task. Due to the large, and diffuse structure throughout the data cube, we utilized the multi-scale clean option with structure scales of [0, 6, 10, 30] pixels, with 1 pixel being equivalent to 0.18''. Multiscale cleaning has been found to be the best method for cleaning diffuse, extended emission at high angular resolution\footnote{\url{https://casaguides.nrao.edu/index.php/ALMA2014_LBC_SVDATA}}. These scales were chosen through trial and error as those that revealed the most emission in regions of interest. CO (2-1) line emission was found in 85 binned 5 km s$^{-1}$ channels, ranging from $25-450$ km s$^{-1}$.  The data were cleaned down to a pre-cleaning RMS level of $\sim$7.5 mJy beam$^{-1}$ per channel, the resulting RMS in 5 km s$^{-1}$ channels averaged $\sim$ 7 mJy beam$^{-1}$. We used the IDL mommaps routine created by Tony Wong\footnote{\url{https://github.com/tonywong94/idl\_mommaps}} to make moment 0, 1, and 2 maps with a local background threshold of 1.5 sigma (shown in Figure \ref{fig:CO21_mirror}). From the moment 0 map, we measured a total CO(2-1) flux (of all flux above 1.5 sigma significance) of $1406 \pm 52$ Jy km s$^{-1}$, remarkably consistent with the flux measured with the SMA by \citet{Lee+17}, $1401 \pm 12$ Jy km s$^{-1}$. We also detect gas to further radial extent, as well as some fainter features that they do not detect, including: the large, stripped eastern plume, decoupled clouds below the plume, and filaments protruding into the southern stripped region (shown and described in Section 5).

\section{Observational Results}

\begin{figure*}
	\plotone{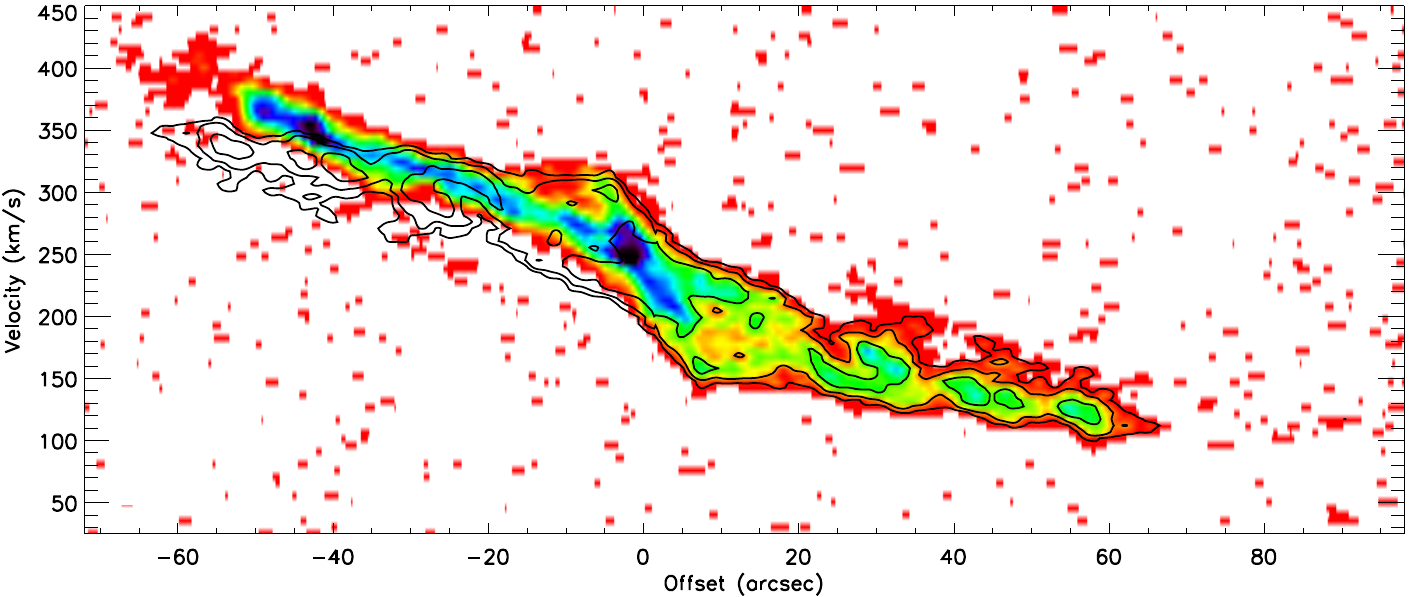}
	\caption{A position velocity diagram (PVD) of CO(2-1) emission along the major axis of NGC 4402, encompassing the full body of the galaxy (the PVD width is indicated in Figure \ref{fig:CO21_mirror}c). In contours a mirror of the right (west) side of the galaxy is overlayed on the left (east) side. The PVD clearly shows the asymmetry between the two sides of the galaxy, particularly past $\sim$40''.}
	\label{fig:PVD_mirror}
\end{figure*}

\subsection{Overall CO distribution}

Our ALMA observations  reveal the molecular gas distribution in the central few kpc of the galaxy NGC 4402. The CO(2-1) moment 0 map in Figure \ref{fig:CO21_mirror}b, c shows that the central kpc has three CO peaks of similar brightness. One of these is close to the nucleus, the other two symmetrically straddle the nucleus, resembling the twin peaks CO distribution commonly observed in the center of barred galaxies \citep{Kenney+92}. In addition, the kinematic minor axis (Figure \ref{fig:CO21_mirror}d) is tilted with respect to the photometric minor axis within $r=0.4$ kpc, as is typically found in barred galaxies. The CO intensity map (Figure \ref{fig:CO21_mirror}b, c) appears to show a strong spiral arm in the west which aligns with the stellar pattern seen at 3.6 $\mu$m in Figure \ref{fig:SF_tracers}, and a significantly weaker one in the east. The CO intensity drops off sharply at a radius of $r=60$''$=4.8$ kpc in the east, and $r=65$''$=5.2$ kpc in the west, although some weak CO features are detected beyond this in the west out to 95''.

The CO distribution shows evidence for ram pressure acting from the southeast, consistent with the estimated projected ram pressure wind direction of $\sim$-45$\degree$ for NGC 4402, based on dust filament orientation and the radio continuum morphology \citep{Crowl+05, Murphy+09, Abramson+14}. The CO distribution is more compressed to the south than the north, as was also noted by \citet{Lee+17}. We also detect a (stripped) extraplanar CO plume at the eastern (leading side) edge of the gas disk in the new ALMA data, a feature first noted in the dust by \citet{Abramson+14}, but not detected in molecular gas in the shallower SMA data presented by \citet{Lee+17}. Beyond $r=55$'' in the east, all of the CO emission is part of the plume and on the north side of the major axis.

\subsection{Overall CO kinematics}

Comparison of the two sides of the galaxy reveals clear evidence of disturbed kinematics. In an undisturbed galaxy, both the moment 1 velocity map, and the PVD along the major axis, should appear symmetric about the minor axis, apart from any bar or spiral arm streaming motions. In NGC 4402 this is not the case, as is highlighted in the position-velocity diagram (PVD) shown in Figure \ref{fig:PVD_mirror}. The PVD  we present encompasses the entire north-south extent of the CO (2-1) in the galaxy (width of $\sim$33''), mirrored about the central velocity of the galaxy, 230 km s$^{-1}$. In contours, we show the west side of the galaxy mirrored onto the east side, to highlight the asymmetric kinematics. The upper envelope of the velocity on the two sides of the galaxy looks relatively symmetric inside $r=25$'', including both the steeply rising circumnuclear component inside $r=5$'', and the region beyond, from $r=5-25$''. Thus, we are confident that the central velocity we chose for folding is correct. 

There are notable differences in the two sides of the galaxy beyond $r=25$'', consistent with the direct pushing of molecular gas by ram pressure. On the east side of the galaxy, upon which the ram pressure would be strongest, the isovelocity contours in the moment 1 map of the galaxy (Figure \ref{fig:CO21_mirror}d) appear to straighten (constant RA) from $r=25-55$'' as opposed to the more typical ``V-shaped" isovelocity contours on the west side (this was also noted with the SMA by \cite{Lee+17}). These V-shaped isovelocity contours are what one would expect to observe in an undisturbed galaxy dominated by rotational motion. In addition, the spiral arm density waves appear to be stronger in the west, as observed in the total CO intensity map and the isovelocity contour map, Figure \ref{fig:CO21_mirror}c \& d. The east side appears to have overall lower velocity dispersion than the west side (Figure \ref{fig:CO21_mirror}e). The PVD in Figure \ref{fig:PVD_mirror} shows more prominent low velocity components near peaks in the spiral arm, suggesting that the lower velocity components may be associated with spiral arm streaming motions, although it may also be ram pressure accelerated gas. Some of the outermost gas ($r > 60$'') in the west is clearly morphologically and kinematically disturbed, probably from ram pressure, and there are other interesting features in the west, but these will be discussed in detail in Paper II.

The two sides of the galaxy begin to clearly diverge in their average velocity around $r=30$''. The velocities of the upper envelope of the PVD on the western side continue to gradually increase with the same slope as $r=10-25$'', and the isovelocity contours of the moment 1 map are consistent with a pattern of mostly circular motions. The velocities on the eastern side rise much more sharply, and the isovelocity pattern shows clear non-circular motions, as shown in Figure \ref{fig:CO21_mirror}d. At $r=55$'', the CO surface brightness peaks at its highest value outside of the circumnuclear region, 0.1 Jy km s$^{-1}$ ($\sim$75 M$_{\odot}$ pc$^{-2}$). The combination of disturbed kinematics in the direction of the ram pressure and high gas surface density strongly suggests compression of this region of the disk due to the ram pressure. We discuss the evidence for this and implications for the ISM evolution in this region in section 6.3.

Beyond $r=60$'' on the east side we observe the most extreme velocity divergence relative to the less disturbed western side of the galaxy. This is in the region of the dust plume, shown in Figure \ref{fig:CO21_zoom}. The plume gas has a significant increase in velocity with radial distance from the galaxy center. At the north-eastern end of the plume, the gas reaches its most extreme velocity, $\sim$60 km s$^{-1}$ higher than any other molecular gas in the galaxy. Ram pressure is the only plausible explanation which could account for both the morphology and kinematics of this feature. The stellar body of the galaxy is symmetric and regular, with no suggestion of any gravitational disturbance \citep{Crowl+05} such as would be expected from a tidal interaction. Furthermore, the effects of ram pressure are expected to be stronger on the eastern side, since this is the leading side of the ram pressure interaction. The plume gas is redshifted with respect to the galaxy, which is the velocity offset expected for stripped gas, as the galaxy is moving towards us (blueshifted) with respect to the cluster.

\section{Velocity Modeling}

We generate a model of a galaxy like NGC 4402 but undisturbed by ram pressure. For this purpose, we used the TiRiFic software package \citep{Jozsa+07} which produces a kinematic model based on circular velocities. TiRiFiC uses a tilted ring parameterization of a rotating disk to create a model data cube that can be smoothed to the same resolution and beam size as our ALMA data. To make this model cube, TiRiFiC requires three main inputs at the radius of each ring: the circular velocity, the velocity dispersion, and the surface brightness.

To measure the circular velocity, we assume that the west side of the galaxy, being more shielded from ram pressure and having a more regular moment 1 map velocity profile, is composed purely of circular and bar motions. We then mirror the west side about the nucleus, to make a map of an entire undisturbed galaxy. To determine the point about which to mirror the west side of the galaxy, we measured V$_{\mathrm{sys}}$ of NCG 4402 using the inner $r \sim 10$'' as an input for DiskFit, a program that uses a tilted ring model to fit the photometry and kinematics of disk galaxies and is able to account for the beam size of the observations (see \cite{Sellwood+15} for documentation). We do not attempt to fit the kinematics of the spiral arms in the galaxy, however, in the central $r \sim 10$'', we do use DiskFit to fit an additional non-circular component of the velocity to account for bar streaming motions. We measure, after 1,000 bootstrapped iterations of DiskFit to reach the minimized chi-squared value of the fit, the center of the galaxy to be RA= 12h26m07.67s DEC= +13\degree06'45.30'' in ICRS J2000. For comparison, the center of the stellar component of the galaxy as measured from the Spitzer 3.6 $\mu$m image is very close, RA= 12h26m07.54s DEC= +13\degree06'45.20'', an offset of 1.8'' (140 pc). The central LSRK velocity is measured by DiskFit to be Vsys = 232 $\pm$ 3 km s$^{-1}$. This velocity is consistent with the central velocity of 230 km s$^{-1}$ chosen by eye to produce a symmetric upper envelope of the velocities on the two sides of the galaxy for $r<25$'', about which we folded the kinematic data for the mirrored PVD in Figure \ref{fig:PVD_mirror}.

Having folded the galaxy about the kinematic and spatial center, we now have an approximation of the velocity field and velocity dispersion field of NGC 4402 if it were undisturbed. We then executed DiskFit on this full, mirrored map, once more fitting for a bar and ignoring non-circular motions from the spiral arms, to measure the circular velocity at the radii of a series of concentric rings.

\begin{figure*}
	\plotone{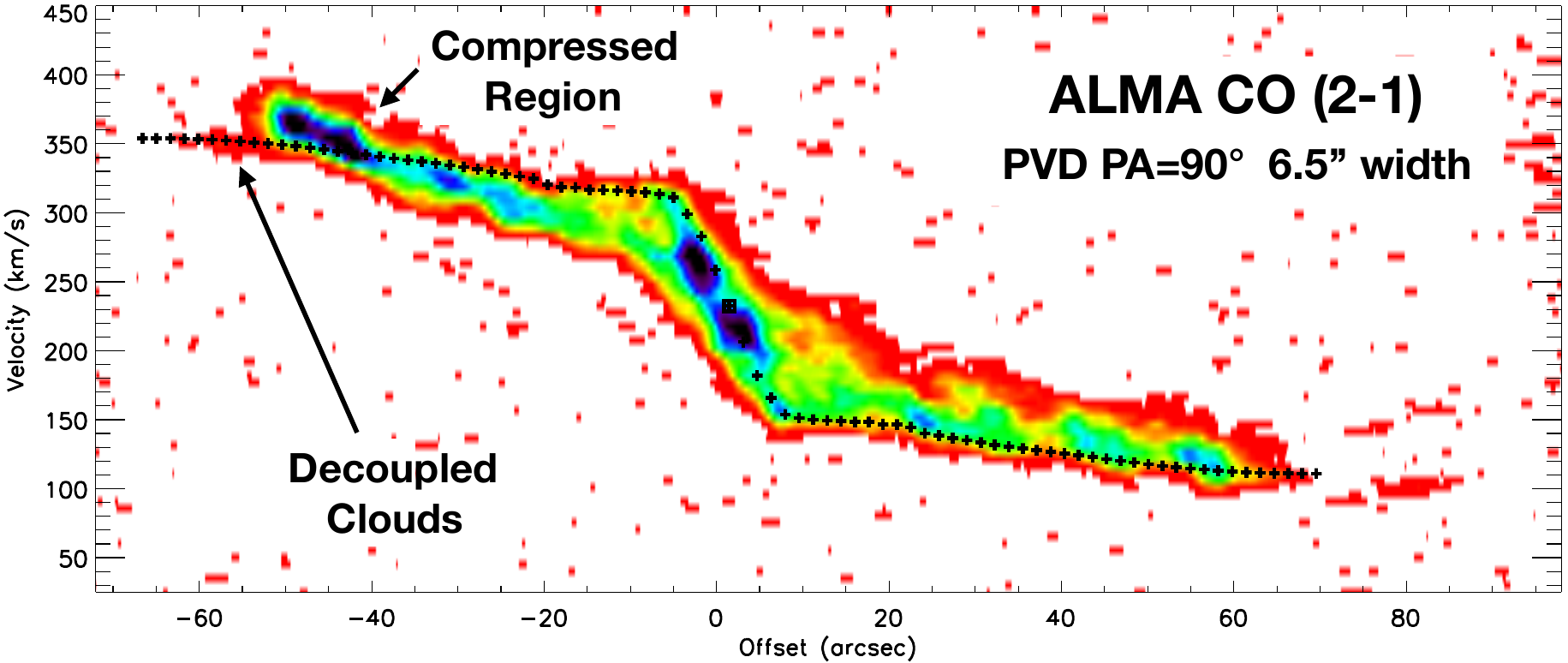}
	\caption{In colors is a PVD slice about the nucleus with a width of 6.5'', relatively thin so as to show circular motions, although some signature from the bar and spiral arms is still present. See the width of the PVD indicated on the moment 0 map (Figure \ref{fig:CO21_mirror}c). Overplotted in points is the rotation curve derived from our model of the kinematics of the western (less disturbed) side of the galaxy, mirrored about the midpoint so that it is also shown plotted along the eastern side.}
	\label{fig:PVD_overplot}
\end{figure*}

The resulting rotation curve is shown plotted on a PVD of the full galaxy in Figure \ref{fig:PVD_overplot}. We note that the velocity of the clouds on the far eastern side of the galaxy with lower velocities than the surrounding plume gas (labeled 'decoupled clouds') is accurately predicted by our model rotation curve, suggesting the velocity of these decoupled clouds is purely due to circular motion, and thus that they have not been accelerated by ram pressure. The properties of these clouds are discussed in more detail in Section 5.

To measure the surface brightness profile, we also used DiskFit on the moment 0 map to measure the surface brightness at each ring. For the final input, the velocity dispersion, we approximated the velocity dispersion profile of the west side of the galaxy, based on the CO (2-1) moment 2 map (Figure \ref{fig:CO21_mirror}e) with a spline fit, to smooth over any discontinuities.

\begin{figure*}
	\plotone{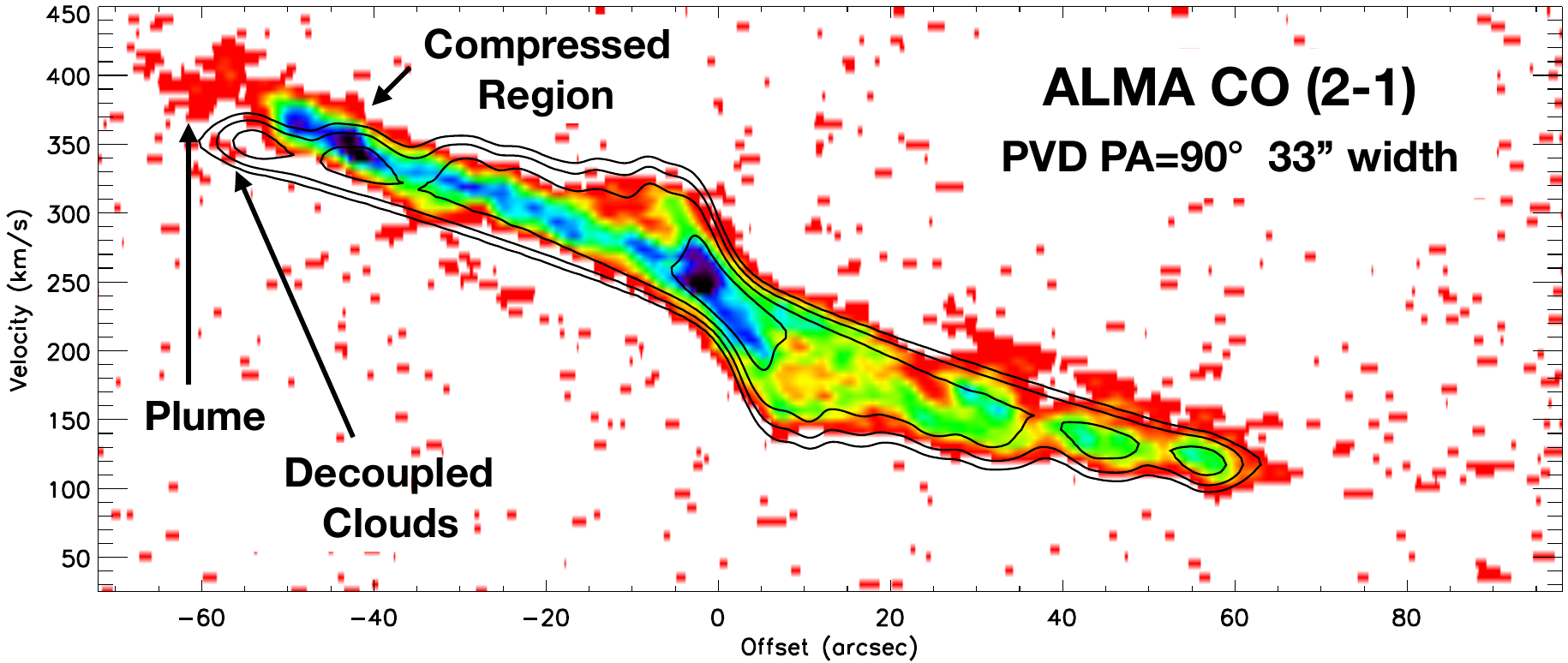}
	\caption{PVD of the full galaxy (with a width of 32.5'') showing the CO(2-1) ALMA data cube, with the circular velocity plus bar motion model cube from Tirific shown in contours, with the lowest contour level of the model corresponding to the weakest emission in the data cube (1.5$\sigma$). See the width of the PVD indicated on the moment 0 map (Figure \ref{fig:CO21_mirror}c).}
	\label{fig:model_pvd}
\end{figure*}

\begin{figure*}
	\plotone{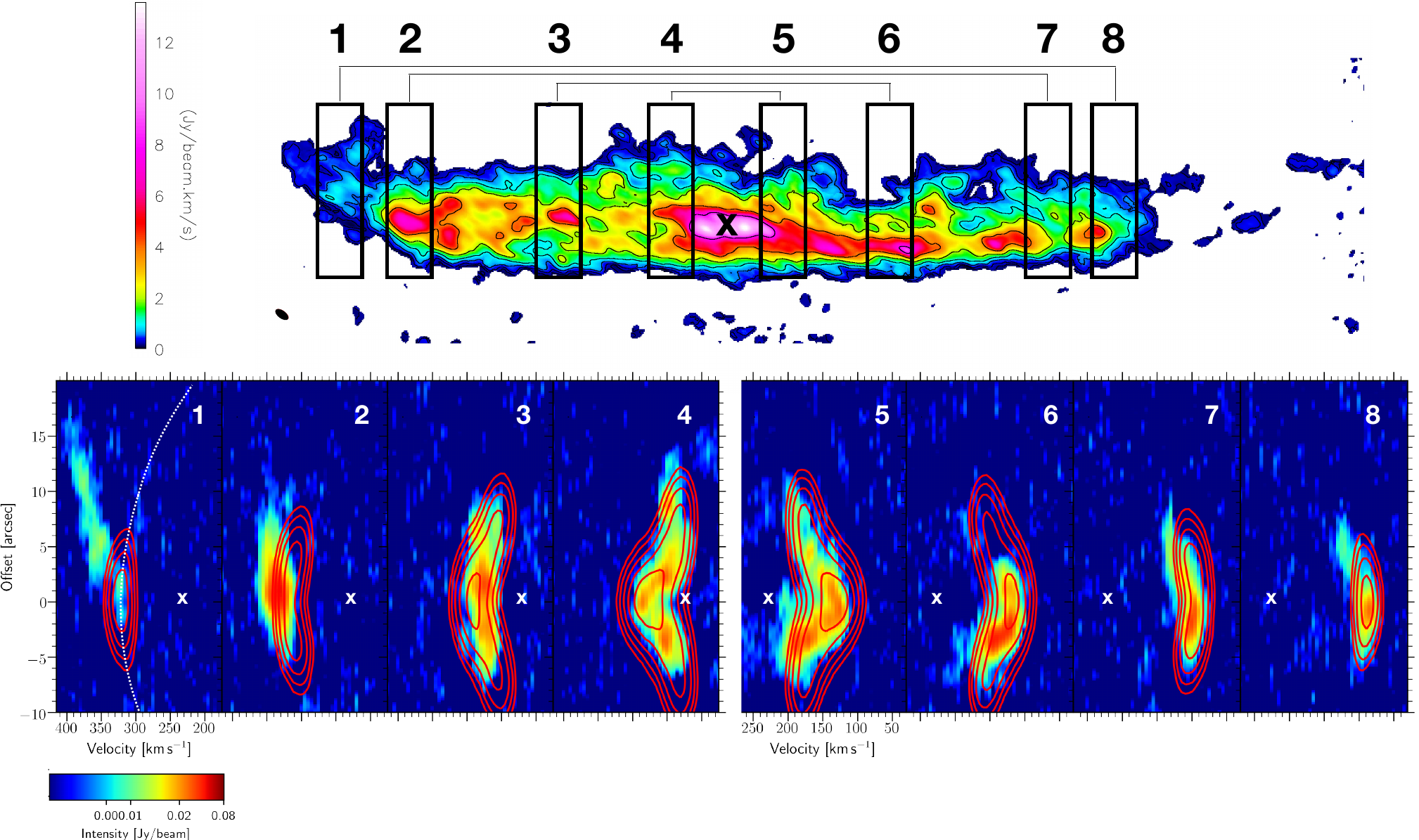}
	\caption{PVD slices, each with a width of 8'' along the minor axis. The x-axis shows the velocity, and y-axis the offset from the center of the galaxy which is marked with an 'x'. Surfaces show the data, while contours show the corresponding PVD slices of the model cube from TiRiFiC, with contour levels of 1$\sigma$, 2$\sigma$, 4$\sigma$, 8$\sigma$, 16$\sigma$, where $\sigma=4.5 \times 10^{-3}$ Jy beam$^{-1}$. Note that on the right side of the galaxy, there is relatively good agreement between the model and data, while on the left side of the galaxy there is a significant offset in both position and velocity. In panel 1, we mark with a white line the projected circular velocity based on the model circular velocity in the disk plane.}
	\label{fig:pvd_grid}
\end{figure*}

The radial profiles of the circular velocity, surface brightness, and velocity dispersion, are then used as inputs for the cube modeling software TiRiFiC. To determine the differences between the two sides of the galaxy, western and eastern, the model we generated is axisymmetric about the nucleus for all three parameters, except for the central $r \leq 10$'' bar region. A PVD comparison of the data and the resulting model cube is shown in Figure \ref{fig:model_pvd}, with the data in colors and the model in contours. It can be seen that around $r=50$'' in the east the data has consistently higher velocities than what the model would predict from an undisturbed rotation curve. This result is very similar to that from the folded PVD shown in Figure \ref{fig:PVD_mirror}. If we were to assume that the gas in the plume originated at its current offset from the disk, i.e. was not accelerated from closer to the disk midplane, we find a velocity divergence from the model projected velocity (indicated by the dotted line in panel 1 of Figure \ref{fig:pvd_grid}) of as much as $\sim$100 km s$^{-1}$. This supports the claim that the velocities in the plume are the result of an outside pressure. The increase in non-circular velocities rises with distance along the plume, which is a signature of ram pressure, which acts from the outside-in, and thus has accelerated outer gas for a longer time. 

This can also be seen in Figure \ref{fig:pvd_grid}, where we show PVD slices of select regions of the data and model, equidistant from the galaxy center. In this figure, we choose regions on both the leading and trailing side at varying radii to show the differences in the two sides of the galaxy in PV space. In panels 3 \& 6, and 4 \& 5, it can be seen that the galaxy and model match up well in PV space on both the leading and trailing side. In panel 2 the $\sim$20 km s$^{-1}$ offset of the compressed region can be clearly seen, and the kinematics of this region looks very different from the region at equivalent distance on the trailing side in panel 7. Finally, the clearly divergent kinematics of the plume gas on the leading side can be seen in panel 1 compared to the trailing side shown on panel 8. There also appears to be some kinematically offset gas in the direction of ram pressure above the disk shown in both panel 7 \& 8, which could be due to either ram pressure weaker than that incident on the leading side of the galaxy, or strong spiral arm streaming motions in these regions.

\section{The Extraplanar Plume, Decoupled Clouds, \& Filaments}

The morphology and kinematics of CO features in the outer eastern region of the galaxy are unlike those previously seen in other ram pressure stripped galaxies. This region of NGC 4402 was previously studied using dust extinction measurements with HST in \cite{Abramson+14}, in which the authors pointed out that this region at the edge of the detectable HI gas disk has spatially decoupled dust clouds where the surrounding gas has already been stripped. 

\begin{figure*}
	\plotone{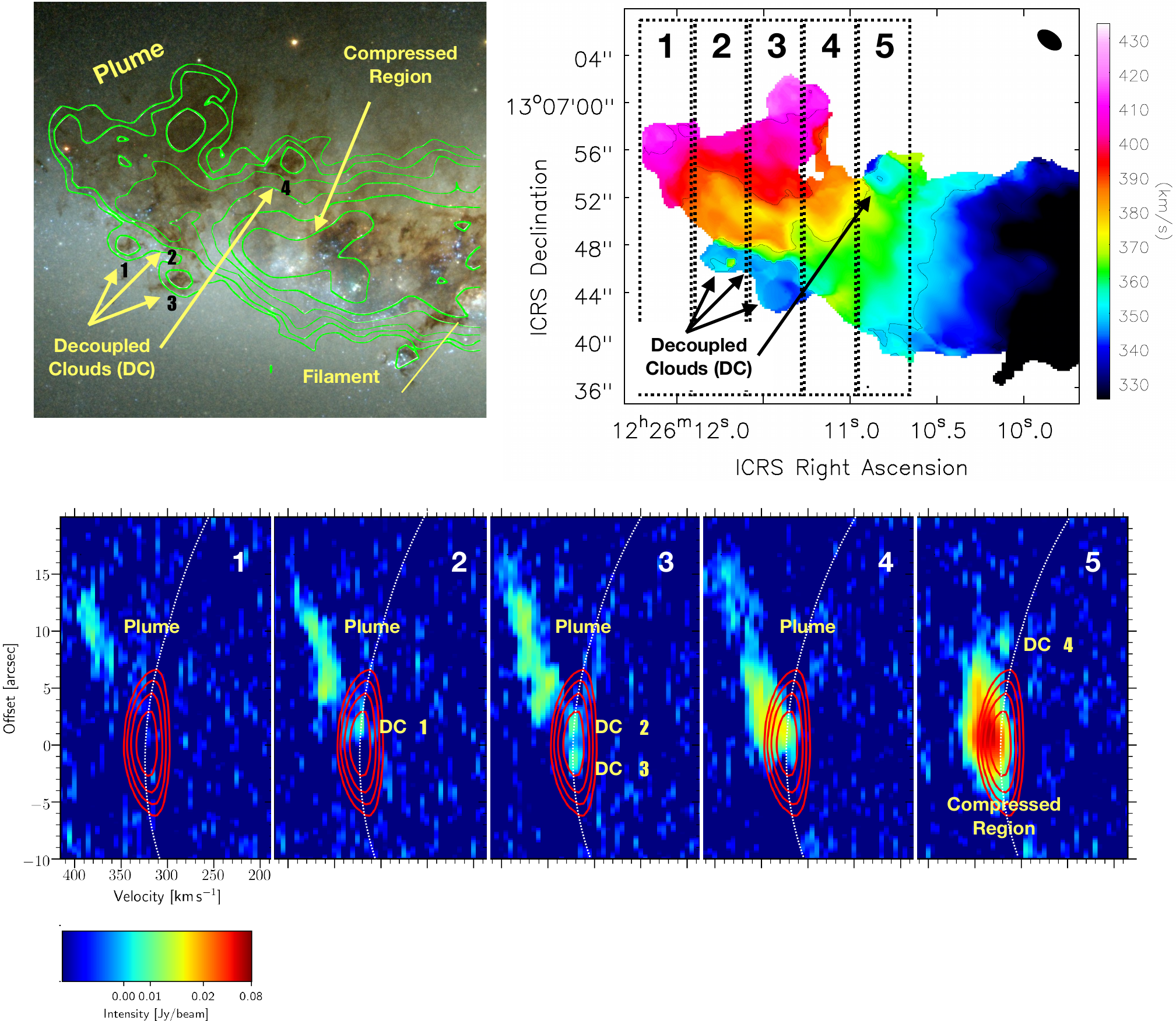}
	\caption{\textbf{Top left (a)}: HST three color image, with CO moment 0 contours, with levels ranging from 0.12 to 7.6 Jy beam$^{-1}$ km s$^{-1}$, with contour increments increasing by a factor of two between each level. \textbf{Top right (b)}: A moment 1 velocity map of the CO with contours ranging from $330-450$ km s$^{-1}$ by increments of 20 km s$^{-1}$. The beam is shown in the upper right corner. \textbf{Bottom (c)}: PVD slices along the minor axis, with the extent of each slice indicated by a box drawn on panel 2. On the x axis is velocity, and on the y the vertical offset from the galaxy center. In red contours is a model of the circular velocity if the galaxy were symmetric about the center (the model is described in section 3.3). The dotted white line in each panel shows the projected circular velocity extended to larger galactocentric radii.}
	\label{fig:CO21_zoom}
\end{figure*}

Indeed, we detect three of these clouds (labelled DC 1, 2, 3) in CO, as shown in Figure \ref{fig:CO21_zoom}a, and confirm that they are kinematically decoupled from the surrounding plume gas, as shown in Figures \ref{fig:CO21_zoom}b \& \ref{fig:CO21_zoom}c  (Panels 2 and 3). These clouds have velocities consistent with normal galactic rotation: the PVDs in (Figure \ref{fig:PVD_mirror}, and Figure \ref{fig:pvd_grid} panel 1), show that the velocities of these decoupled clouds match those of gas on the west side of the galaxy at the same radius. The small clouds have apparently decoupled from the gas in the plume, which is further offset from the major axis, and offset in velocity by $20-75$ km s$^{-1}$ with respect to the predicted rotation velocities. These properties are all consistent with the plume gas being accelerated from the disk midplane by ram pressure, while the DCs have resisted this acceleration.

Why has gas in the decoupled clouds resisted stripping, while the surrounding plume gas is lifted away from the disk? One possible explanation that we investigate here and in Section 6.1 is whether the decoupled clouds are more dense than the surrounding ISM, and thus more resistant to stripping.

In order to estimate the surface density of these features, first, we derive the flux of each feature from the moment 0 map (Figure \ref{fig:CO21_mirror}). Following the conversions laid out in \citet{Solomon+97, Lee+17} we then convert the total integrated flux to a line luminosity $L_{\mathrm{CO}}^{\prime}(2-1)$:

$$L_{\mathrm{CO}}^{\prime}(2-1)=3.25 \times 10^{7} S_{\mathrm{CO}} v_{\mathrm{obs}}^{-2} D_{L}^{2}(1+z)^{-3}$$

where S$_{\mathrm{CO}}$ is the total integrated CO (2-1) flux in Jy km s$^{-1}$, $v_{\mathrm{obs}}$ is the frequency of the CO (2-1) transition in GHz, $D_{L}$ is the luminosity distance of NGC 4402 (17 Mpc), and $z$ is the redshift of the galaxy (0.0008).

To then convert from CO (2-1) luminosity to total mass of H$_2$, we use the following equation:

$$
M_{\mathrm{H}_{2}}=1.34\frac{\alpha_{\mathrm{CO}}}{R_{21}} L_{\mathrm{CO}}^{\prime}(2-1)
$$

where we adopt the CO-H$_2$ relation based on that measured for the Milky Way disk, $\alpha=4.3$ $\mathrm{M}_{\odot} \,\mathrm{pc}^{-2} \, (\mathrm{K} \, \mathrm{km} \, \mathrm{s}^{-1})^{-1}$ \citep{Bolatto+13}, and the CO (2-1)/(1-0) ratio of nearby galaxies, $R_{21}\approx 0.8$ from \citet{Leroy+09}; to account for the mass of helium, we multiply $\Sigma_{\mathrm{H_2}}$ by 1.34 \citep{Klessen+16}. We note that while we assume a uniform $X_{\mathrm{CO}}$ throughout, it is certainly possible that it could vary across the region. To estimate the surface area of each feature, we measure the size of the feature from dust extinction in HST images, and assume that the CO and dust are coextensive. Since the resolution of our HST ($\sim$0.1'') data is much better than that of our ALMA data ($\sim$1 $\times$ 2''), this is probably a more precise method of measuring the physical extent of these features. Due to the narrow line widths ($\sim$15 km s$^{-1}$) and simple line profiles of the decoupled clouds, they are likely single clouds. Our results for the cloud properties are shown in Table 1.

\begin{table*}
\label{tab:main_table}
\movetableright=-0.5in
\begin{tabular}{cccccccc}
\hline
\head{1.5cm}{\textbf{Feature}} & \head{2.0cm}{\textbf{Integrated Flux (Jy km s$^{-1}$)}} & \head{1.0cm}{\textbf{Mass M$_{\mathrm{H_2}+\mathrm{He}}$ (M$_{\odot}$)}} & \head{2.0cm}{\textbf{Measured mass M$_{\mathrm{H_2}+\mathrm{He}}$ from dust (M$_{\odot}$)}} & \head{1.5cm}{\textbf{Area (pc$^2$)}} & \head{2.0cm}{\textbf{Surface Density (M$_{\odot}$ pc$^{-2}$)}} & \head{1.5cm}{\textbf{Line Width (km s$^{-1}$)}} & \head{2.0cm}{\textbf{Spectral Surface Density (M$_{\odot}$ pc$^{-2}$ km$^{-1}$ s)}} \\ \hline
\textbf{Cloud 1} & 0.53 & 6.8 $\times$ 10$^5$ & 2.1 $\times$ 10$^5$ & 2.7 $\times$ 10$^4$ & 26 & 22 & 1.2\\
\textbf{Cloud 2} & 0.53 & 6.8 $\times$ 10$^5$ & - & 4.1 $\times$ 10$^4$ & 16 & 13 & 1.2 \\
\textbf{Cloud 3} & 0.64 & 8.1 $\times$ 10$^5$ & 5.4 $\times$ 10$^5$ & 5.7 $\times$ 10$^4$ & 14 & 12 & 1.12 \\
\textbf{Cloud 4} & 0.99 & 1.3 $\times$ 10$^6$ & - & 4.6 $\times$ 10$^4$ & 28 & 14 & 2.0 \\
\textbf{Filament} & 0.41 & 5.2 $\times$ 10$^5$ & 8.9 $\times$ 10$^5$ & 7.5 $\times$ 10$^4$ & 7.0 & 22 & 0.32 \\
\textbf{Plume} & 21.4 & 2.7 $\times$ 10$^7$ & - & 1.1 $\times$ 10$^6$ & 25 & 42 & 0.60 \\
\end{tabular}
\caption{Properties of selected CO features in NGC 4402. From left to right, in column (1) the feature name. (2): The integrated CO(2-1) flux based on the moment 0 map. (3): Total mass of H$_2$ \& He assuming $\alpha=4.3 \, \mathrm{M}_{\odot} \,\mathrm{pc}^{-2} \, (\mathrm{K} \, \mathrm{km} \, \mathrm{s}^{-1})^{-1}$ and $R_{21} \approx 0.8$, and the mass fraction of He/H$_2$ to be 1.34. (4) Minimum mass of M$_{\mathrm{H_2}+\mathrm{He}}$ estimated from the dust extinction $A_V$ from \citet{Abramson+14} for clouds also found in that paper. We note this included both HI and H$_2$, whereas our mass estimate in column (3) does not include HI. (5) Size of features based on spatial extent of the dust clouds, from \citet{Abramson+14}. (6): Estimated average surface density of these features by dividing column (3) by column (6). (7) The line width $\sigma$ of the features, estimated from the FWHM of a Gaussian fit to the line profile of each feature. (8) The ``spectral'' surface density of each feature, measured by dividing column (6) by column (7).}
\end{table*}

We find relatively good agreement between our estimates of the gas mass M$_{\mathrm{H_2}+\mathrm{He}}$ from CO flux of the small features, decoupled clouds, and filaments, and estimates of the mass from dust extinction of clouds measured by \citet{Abramson+14}. In \citet{Abramson+14}, the authors stated that their estimates are likely lower limits on the total mass, since dust column density can exceed the estimate from optical extinction depending on the three dimensional distribution of the dust and stars. They thus suggested the total mass of the ISM features could be about a factor of ten higher, bringing the decoupled clouds in line with the surface densities of GMC-like clouds measured in other studies. We find, however, that the mass estimated from dust extinction is roughly reliable, and that these decoupled clouds have average surface densities (over $80-160$ pc) of $\sim$ $15-30$ M$_{\odot}$ pc$^{-2}$, approaching the lower end of measured GMC surface densities in the outer regions of the Milky Way $\sim$30 M$_{\odot}$ pc$^{-2}$ for GMCs \citep{Heyer+15}. We further discuss the properties of the decoupled clouds in Section 6.1.

Another likely example of cloud decoupling is the filament, marked in Figure \ref{fig:CO21_zoom}a, first identified in \citet{Crowl+05} from dust extinction, and shown in high resolution with HST in \citet{Abramson+14}. It is approximately 600 pc long, and has a clump of young, blue stars at the head, as seen in the HST color image (see Figure \ref{fig:CO21_mirror}a). The filament sticks out into the otherwise stripped southern zone of the galaxy. It may be supported by magnetic tension \citep{Abramson+16}, as suggested for similar features seen in the RPS galaxy NGC 4921 in the Coma cluster \citep{Kenney+15}.

\section{Discussion}
\subsection{What surface density of gas can be stripped by RP?}

As our data shows strong evidence for the direct stripping of molecular gas from the disk of NGC 4402 via ram pressure, we wish to estimate the strength of ram pressure and the opposing restoring force from the disk around the stripped region. By estimating the ICM gas density $\rho_{I C M}$, the relative velocity of the galaxy with respect to the cluster $v$, and the restoring force per unit mass $d\Phi_g/dz$, we can use the \citet{Gunn+72} stripping criteria $\rho_{I C M} v^{2} \geqslant \Sigma_{I S M} \frac{d \Phi}{d z}$  to estimate the surface density of gas $\Sigma_{I S M}$ that would be stripped, and compare it to our estimates for the surface densities of features in Table 1. The projected distance of NGC 4402 from the cluster center is 0.4 Mpc, and its 3D distance would be larger than this. From measurements of the X-ray surface brightness profile of Virgo from \citet{Schindler+99}, at the 3D cluster radius of 0.4 Mpc, the density is estimated to be $\rho_{\mathrm{ICM}}=3 \times 10^{-28}$ g cm$^{-3}$,  and further out it would be less. However NGC 4402 is located only about 50 kpc in projected distance from the elliptical M86, and may be a member of the M86 subcluster, which has its own local peak in X-ray surface brightness and ICM density.  \citet{Schindler+99} did not model the M86 region, but from comparison of the X-Ray surface brightness near NGC 4402 and at other locations at the same projected radius, we estimate the ICM density near NGC 4402 could be up to two times higher than the above estimate. Thus we adopt an ICM density in the range of $3 \times 10^{-28}$ g cm$^{-3}$, up to a possible maximum of $6 \times 10^{-28}$ g cm$^{-3}$.

The line of sight velocity of NGC 4402 with respect to the cluster is $v_{\rm gal}-v_{\rm Virgo} = 847$ km s$^{-1}$, and the 3D velocity would be larger than this. The orbital velocity of a typical Virgo galaxy at a 3D radius of 0.4 Mpc is predicted to be $\sim$1300 km s$^{-1}$ \citep{Binggeli+93}, although the maximum could be as high as $\sim$1800 km s$^{-1}$ \citep{Vollmer+01a}. Thus we adopt a relative velocity range of $1300-1800$ km s$^{-1}$. If NGC 4402 were not a member of the M86 subcluster, but happened to be falling through it, NGC 4402 could be moving at the higher end of the velocity estimate of 1800 km s$^{-1}$ and experiencing the higher estimated ICM density ($6 \times 10^{-28}$ g cm$^{-3}$) associated with the M86 subcluster. Therefore, in summary, the ram pressure at NGC 4402 is predicted to be in the range of $P_{\mathrm{ram}}=\rho v^2 \sim 5 \times 10^{-12}$ to $2 \times 10^{-11}$ dyne cm$^{-2}$.

We estimate the gravitational restoring force per unit mass $\frac{d \Phi}{d z}$ as approximately equal to $V^2_{rot} R^{-1}$, as done in \citet{Lee+17}. At the radial distance of the plume, $r \sim 60$''$\, = 4.8$ kpc, the circular velocity predicted from our DiskFit model is 119 km s$^{-1}$, so $V^2_{rot} R^{-1} = 9.6 \times 10^{-14}$ km s$^{-2}$. Using the Gunn \& Gott formula, we estimate the surface density of gas that would be stripped under these conditions as $\Sigma_{I S M} = \rho_{I C M} v^{2} V^{-2}_{rot} R$ $\approx$ 2.5 M$_{\odot}$ pc$^{-2}$ for the lowest estimated density and velocity, and $\Sigma_{I S M} \approx 10$ M$_{\odot}$ pc$^{-2}$ for the upper limit.

The small clouds we have labelled as `decoupled clouds' ($1-3$) are located right below the plume, and are spatially and kinematically distinct from the plume, with velocities consistent with normal rotation, and thus, they appear to not be accelerated by ram pressure. In these respects, these clouds are consistent with denser clouds that decouple from surrounding lower density gas that is stripped. With surface densities of $14-28$ M$_{\odot}$ pc$^{-2}$, these clouds have higher surface densities than the surface densities of $2.5-10$ M$_{\odot}$ pc$^{-2}$ expected to be stripped from $r \sim 5$ kpc in NGC 4402. Thus these clouds are consistent with the expectations of clouds that have resisted stripping because of high surface densities.

The observed surface density of the plume gas is similar to that of the decoupled clouds, which at first might seem inconsistent with the observation that the plume gas is being directly stripped.  However the relevant surface density to compare with these theoretical expectations for stripping is the gas surface density in the wind direction. This is different from the observed (plane of sky) surface density, since the ram pressure has a component in the plane of the sky (and perpendicular to the disk plane) as well as along the line of sight.

If we assume that the gas features have a similar depth along the line-of-sight as they do on the sky, then the plume gas would have a depth of $\sim$1 kpc ($\sim$5 beams) and the smaller clouds would have a depth of $\sim$200 pc ($\sim$1 beam), a difference of a factor of $\sim$5. Thus the angular correction factor would be $\sim$5 times larger for the plume than the decoupled clouds, and the surface density of the plume in the wind direction could be $\sim$$2-5$ times less than the observed surface on the sky, depending on the precise wind angle.

Further evidence that the plume gas may be significantly extended along the line-of-sight comes from its large line width compared to the decoupled clouds. While the plume and the decoupled clouds have similar total surface densities, the line widths in the plume are much larger than those of the decoupled clouds, so the surface density per velocity interval, i.e. the `spectral' surface density (Table 1), is $2-4$ times lower in the plume than in the decoupled clouds.

This difference in spectral surface density suggests that the plume gas is more diffuse, with a lower intrinsic \textbf{volume} density, than the decoupled clouds. In the Milky Way galaxy, \citet{Roman+16} find that diffuse molecular gas has a lower spectral surface density, on average than dense, GMC associated gas. Given its large line width, the plume may contain multiple clouds superimposed, such that the surface density in the wind direction for the plume is plausibly $\sim 2-4$ times lower than the observed surface density, or 6-12 M$_{\odot}$ pc$^{-2}$. This would mean the plume has a lower surface density in the wind direction than the decoupled clouds, and is likely consistent with having a surface density in the range of $2.5-10$ M$_{\odot}$ pc$^{-2}$ that could be stripped from $r \sim 5$ kpc in NGC 4402.

We note that our estimates of molecular gas surface densities are based on a standard CO-H$_2$ relationship (X-factor, or X$_{\mathrm{CO}}$), and we have no observational constraints on the CO-H$_2$ relationship in the plume or decoupled clouds of NGC 4402. It will be of interest to measure other molecular lines in the future to constrain X$_{\mathrm{CO}}$.

Our calculation of the maximum density of ISM gas that could be stripped is less than the lower end of the measured density of GMCs in the outer regions of the Milky Way \citep{Heyer+15}, suggesting that ram pressure is not likely to directly strip GMCs. On the other hand, it is probably strong enough to strip more diffuse molecular gas, which can comprise a significant fraction of the total molecular gas in a galaxy. In the Milky Way \citet{Roman+16} estimate that `diffuse' molecular gas, as identified from a non-detection in $^{13}$CO, comprises 25\% of the total molecular gas, although the fraction increases with galactocentric radius up to 50\% at a radius of 15 kpc. Conversely, this means that $\sim$ 50-75\% of the molecular gas is `dense', and likely associated with GMCs. However, we observe that on the leading side of NGC 4402 only 12\% of the CO is in denser decoupled clouds below the plume, that are dense enough they appear to be kinematically unaffected by ram pressure. 88\% of the CO appears to be undergoing stripping in the plume, and we have measured that it has a lower spectral surface density than that of the DCs, and thus is likely `diffuse'. A near 90:10 ratio of diffuse to GMC gas seems incompatible with the fraction of dense gas we would expect in this region based on the findings of \citet{Roman+16}. And furthermore, the presence of significant ongoing star formation in this part of NGC 4402 indicates that much of the molecular gas in this region must have relatively recently been part of a GMC.

\subsection{Timescales for stripped gas}

If ram pressure is too weak to strip GMCs directly, they must be disrupted by some other effect on a timescale shorter than the ram pressure stripping timescale. We estimate this to be the time that the ISM spends on the leading quadrant of the galaxy, where it is not shielded by any intervening ISM, and thus experiences close to the maximum ram pressure. We estimate the time to accelerate gas at the end of the plume as $t= 2*\mathrm{cot}(\theta) * \mathrm{L}_{\mathrm{sky}} / v_{\mathrm{LOS}}$,  where L$_{\mathrm{sky}}$ and v$_{\mathrm{LOS}}$ are the observed projected components of length and velocity, and $\theta$ is the angle of the RP wind with respect to the line of sight (see \citet{Kenney+14} for a derivation of this formula). This assumes constant acceleration starting from zero velocity. The angle of the RP wind with respect to the line of sight is not known, but is probably in the range $30-60$ degrees based on the likely total velocity of galaxies in Virgo \citep{Vollmer+01a}, corresponding to a range of $0.58-1.73$ for the cotangent factor. For $\mathrm{L}_{\mathrm{sky}} = 1$ kpc, $v_{\mathrm{LOS}} = 60$ km s$^{-1}$, and a -45 degree wind-LOS  angle, t = $3.2 \times 10^7$ yr, or 32 Myr. Based on our rotation curve, we estimate the orbital period of NGC 4402 to be $\sim$250 Myr at the radial distance of the plume. Therefore, gas in the plume has likely been experiencing strongly effects of RPS for $\sim$15\% of the orbital time, consistent with gas on the leading side having been exclusively accelerated from the leading side.

For estimating this timescale, we also consider the lower and upper limit, the upper limit being for $\theta = 30$ degrees, $t=5.7 \times 10^7$ yr, and the lower limit, for $\theta = 60$ degrees is $t=1.9 \times 10^7$ yr. Because we detect no clouds with densities like those of GMCs in this region, GMCs must be disrupted, and prevented from reforming, on a timescale below $\sim 20-60$ Myr. One possible explanation for the lack of GMCs detected in the stripped zone is disruption by stellar feedback, then stripping of the resulting disrupted gas by ram pressure preventing the normal reforming via cooling and gravitational collapse of GMC gas. This hypothesis is supported by results from \citet{Kruijssen+19}, in which they find evidence for the dispersal of GMCs in NGC 300 on short timescales of $\sim$1.5 Myr after the formation of massive stars due to photo-ionization and stellar winds, limiting the average lifetime of GMCs to $\sim$10 Myr. 

The compression from ram pressure could increase the fraction of the ISM that becomes GMCs, which then form stars. This may synchronizes and concentrate the formation of GMCs in a similar manner to how spiral arms act on the gas \citep{Vogel+88}. The resulting concentrated star formation and its feedback results in the synchronized destruction of the GMCs, the remnants of which would be less dense than the original GMC, and thus easier to strip.

\subsection{Direct vs indirect stripping}

Throughout this paper we have referred to `direct stripping', by which we mean that gas that was molecular in the disk is stripped as molecular gas and remains molecular on the way out. The alternative to this would be `indirect stripping', by which molecular clouds in the disk, instead of being directly pushed by ram pressure, have their gas revert to a lower density atomic or ionized state as a result of star formation feedback or natural evolution, and then have the gas stripped. In this scenario, molecular gas we detect in the plume would have been formed in situ from this stripped atomic or ionized gas after some period of time. 

Our evidence supports that GMCs are not directly stripped, but that their gas is somehow effectively removed within a short time after the surrounding gas is stripped. But we also find our evidence is consistent with direct stripping of diffuse molecular gas. Most of the molecular gas is gone in the outer radial zone ($4-5$ kpc) on the leading side of the disk, and there is stripped molecular gas a short distance downstream in the plume. The likely surface density of this gas is comparable to that of diffuse molecular gas in spiral galaxies, and is probably low enough to be directly stripped, based on our estimates of the likely ram pressure at NGC 4402 in section 6.1. We cannot prove that this gas began as molecular gas in the disk, but that is the simplest explanation.

If this gas was stripped when it was not molecular, then the molecules form in situ very quickly after stripping. The molecular plume starts within $\sim$2''$=200$ pc of the decoupled clouds. In sec 6.2 we calculate an acceleration time of 32 Myr for gas to reach the end of the plume at $\sim$1 kpc, and this corresponds to an acceleration time of $\sim$6 Myr for the start of the plume (the portion closest to the disk midplane) at $\sim$200 pc. Thus, in situ formation of molecular gas would have to happen on a timescale of a few Myr.

For most purposes related to galaxy evolution, whether or not the gas remains molecular the whole time doesn't matter since most of the molecular gas is gone in this radial zone of the disk, and there is stripped molecular gas a short distance downstream from this zone.

\subsection{Evolutionary stages of ram pressure}
\label{Evolution}

\begin{figure}
	\plotone{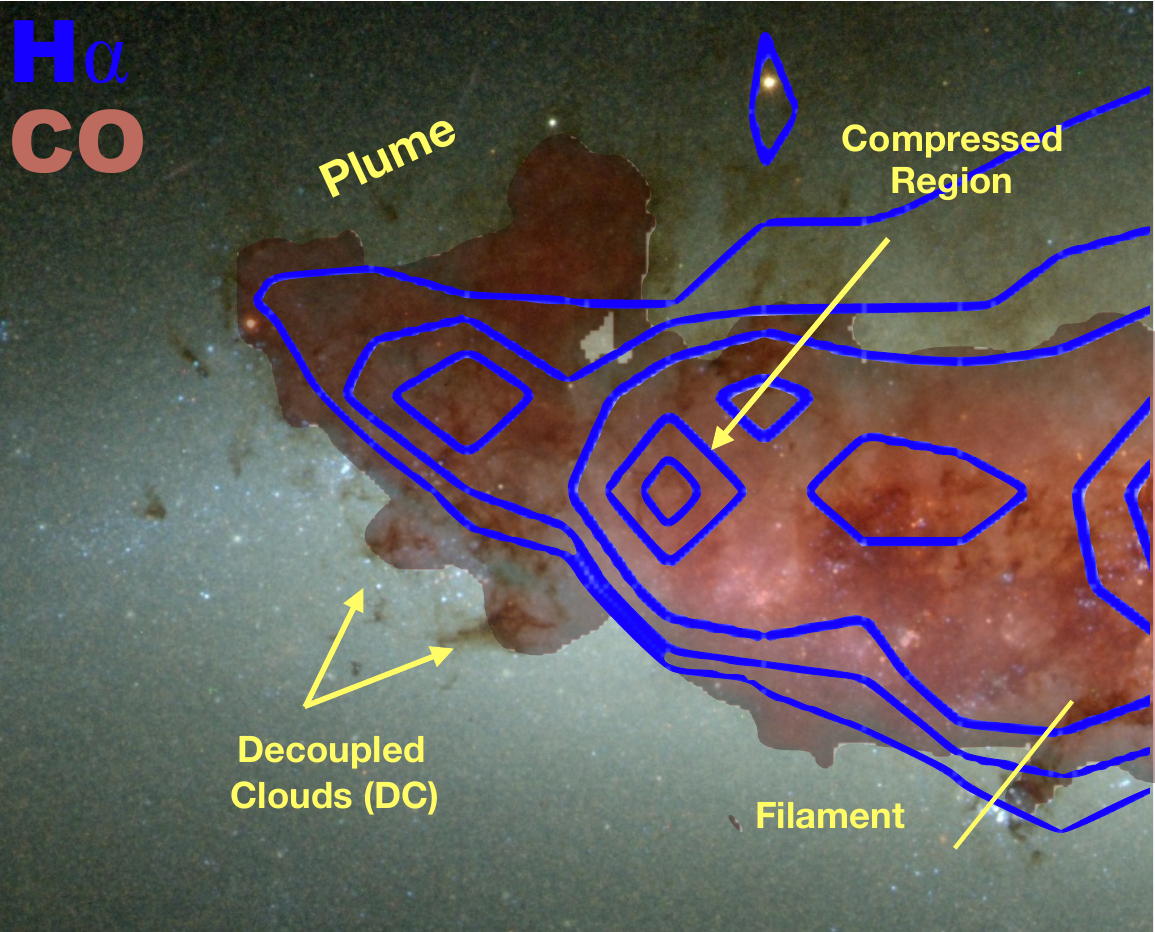}
	\caption{HST three color image, with CO moment 0 shown as a reddish overlay, and H$\alpha$ (from \citet{Kenney+08}) shown with the blue contours. H$\alpha$ contour levels vary from 4 $\times \, 10^{36}$ to 1 $\times \, 10^{38}$ ergs s$^{-1}$ arcsec$^{-2}$ by factors of 2. There is some overlap of the H$\alpha$ and the part of the plume closest to the ram pressure front, which may be the result of star formation triggered by compression here. The decoupled clouds are not detected in H$\alpha$, and thus likely have no ongoing or recent (within the last $\sim$ 10 Myr) star formation.}
	\label{fig:HA_CO}
\end{figure}

Our high resolution ALMA observations show the molecular gas undergoing both compression and stripping as a result of ram pressure. We suggest that these represent two different evolutionary stages of the ISM under ram pressure. The compressed region (see Figure \ref{fig:CO21_zoom}) shows the first stage of major RPS effects. While gas in this region does not reach non-circular velocities near that of the plume, Figure \ref{fig:pvd_grid} shows that it is kinematically offset by $\sim$20 km s$^{-1}$ in the direction of the ram pressure from the predicted rotation velocity at its location. The ISM is compressed which likely results in both direct concentration of molecular gas and/or enhanced formation of molecular gas via compression of the atomic gas, as an increase in relative gas and dust density could lead to more H$_2$ formation on dust grains. This increase is enough to make the center of the compressed region the brightest peak in CO surface brightness outside the circumnuclear region of the galaxy, corresponding to a peak surface density of $\sim$75 M$_{\odot}$ pc$^{-2}$. It is possible this triggers intense star formation in the compressed region, as is seen on the leading side of a number of other galaxies under ram pressure \citep{Koopmann+04, Vollmer+12, Lee+16, Vulcani+18, Roberts+20}. Both FUV and H$\alpha$ observations (Figure \ref{fig:Virgo_cluster}b \& c and Figure \ref{fig:HA_CO}) of the leading side of NGC 4402 show spatially correlated strong peaks in emission. The FUV image shows by far the strongest emission at the leading, eastern edge. The ram pressure wind has probably blown away some of the surrounding gas and dust in this star forming region, so that fewer of the UV photons are absorbed by gas and dust. Furthermore, the removal of gas and dust in this region means that tracers of star formation like H$\alpha$ and FIR may underestimate the star formation rate in this region, as the material that normally emits in this range (gas in HII regions and dust) is removed. Moreover, it is possible that the typical relation of H$\alpha$-SFR may under-predict the true SFR in this region. Strong ram pressure stripping of the gas and dust around HII regions may remove this material which would otherwise absorb some fraction of the Lyman continuum ionizing photons from young massive stars. In the ram pressure stripped jellyfish galaxy JW100, \citet{Poggianti+19c} found that the escape fraction of Lyman continuum photons around HII regions in the tail was as high as $\sim$52\%, while the average escape fraction estimated for HII regions in the GASP sample was $\sim$18\%. We believe it is likely, given the FUV peak and the local peaks in H$\alpha$ and 24 $\mu$m, that the leading side of the galaxy has recently undergone strong star formation.

We find evidence in the plume region for an earlier evolutionary stage of compressed gas and triggered star formation. Figure \ref{fig:HA_CO} shows contours of H$\alpha$ emission superimposed on the HST and ALMA CO(2-1) images. Within the bottom part of the plume, there is an H$\alpha$ ridge that angles upward from the disk and is offset from the major axis in the direction of ram pressure. There are a number of visible blue stars in the region of the disk midplane cleared of dust just south of the H$\alpha$ ridge, indicating recent star formation. The gas that formed the young stars powering the HII regions in the plume detected in H$\alpha$ was likely displaced from the disk midplane by ram pressure. It is likely that the compressed region, indicated in Figure \ref{fig:HA_CO}, is at an earlier evolutionary stage of this process. From the PVD shown in Figure \ref{fig:pvd_grid} we can see that the compressed region has been kinematically disturbed by ram pressure, but there has not yet been sufficient time and acceleration for it to have caused a significant spatial offset. In several more Myr, the acceleration this gas has received will likely cause it to be spatially offset from the disk midplane, like the offset observed for the HII regions in the plume region. Thus the stripped zone appears to be consistent with being the expected later evolutionary stage of the compressed zone.

In regions of intense star formation, triggered by ram pressure compression, stellar feedback could have a significant effect on local stripping efficiency. Stellar feedback would blow gas out from the region, both accelerating it towards the escape velocity ($\sim$300 km s$^{-1}$ at this radius for NGC 4402) and lowering the density of the ISM, and thus requiring less ram pressure to eventually strip the gas. Stellar feedback may also lead to the formation of holes in the ISM, allowing RPS to ablate the densest regions \citep{Quilis+00}. In the high resolution gas simulations by \citet{Tonnesen+10, Tonnesen+12}, the authors found ram pressure of $\mathrm{P}_{\mathrm{ram}}=6.4 \times 10^{-12}$ dyne cm$^{-2}$ (very similar to our estimate of the ram pressure experienced by NGC 4402) was too weak to strip gas with the density of typical GMCs directly, but did find gradual stripping via hydrodynamical ablation of the envelopes surrounding the dense, GMC-like clumps. However, the authors notably did not include stellar feedback in these simulations. As we find that any GMCs which avoid the initial stripping do not survive for long, the inclusion of proper prescriptions for stellar feedback in simulations may be key to accurate predictions for the fate of GMCs during stripping events.

\section{Summary}

We have presented new ALMA CO(2-1) observations of one of the nearest and best examples of a galaxy experiencing ram pressure stripping, NGC 4402 in the Virgo cluster. We have found clear evidence from both the morphology and kinematics for both the compression and direct stripping of molecular gas via ram pressure.

\begin{description}
	\item[$\bullet$ Direct stripping of molecular gas] We have detected a large (width of $\sim$1 kpc, height of $\sim$2 kpc, and mass of M$_{\mathrm{H}_2+\mathrm{He}}=2 \times 10^7$ M$_{\odot}$) extraplanar plume of molecular gas on the eastern, leading side of the galaxy at $r \sim 4-5$ kpc, first seen in HST observations of dust extinction. We show that the plume gas has non-circular motions that are likely the result of ram pressure acceleration. At the north-eastern end of the plume, the gas reaches its most extreme velocity, $\sim$100 km s$^{-1}$ higher than the predicted projected circular velocity at this location, and $\sim$60 km s$^{-1}$ higher than gas in the nearby disk from where it probably originated. The plume is both spatially and kinematically offset in the direction consistent with the expected ram pressure wind direction.
	
	\item[$\bullet$ Decoupled clouds] Just below the extraplanar plume of gas, we find isolated clouds that have kinematics consistent with simple rotation, i.e. no disturbance from ram pressure. This suggests that some of the material on the leading side has resisted acceleration by ram pressure. The detected decoupled clouds represent only $\sim$12\% of the total mass of molecular gas in this zone. At the resolution of our ALMA data, $\sim1-2$''$\, =80-160$ pc, the observed surface brightness of CO in the plume and decoupled clouds are similar, with estimated surface densities of molecular gas of $\sim15-30$ M$_{\odot}$ pc$^{-2}$. The plume gas has a significantly lower spectral surface density than the decoupled clouds, suggesting that it is more diffuse and spread over a much larger volume than the decoupled clouds. We claim that decoupled clouds have resisted stripping since they have measured gas surface densities which exceed the estimated surface density of gas that can be stripped by our estimates of the predicted strength of ram pressure.
	
	\item[$\bullet$ Compression of molecular gas] Just inside of the plume of stripped material, closer to the galactic center at $r=3.5$ kpc, we find the highest peak in the CO surface brightness outside of the circumnuclear region. While this region has no visible plume, and no gas has velocity offsets as high as the stripped material in the plume, this region does appear to be kinematically offset by $\sim$20 km s$^{-1}$ from the expected rotation velocity in the direction consistent with the ram pressure, indicating acceleration from ram pressure. This suggests this is a zone of molecular gas compressed by ram pressure. The large FUV and H$\alpha$ peaks in this region indicate this is a region undergoing intense star formation.
	
	\item[$\bullet$ Evolutionary stages of the RP interaction] We propose that the compressed region at $r$ = 3.5 kpc and the plume plus decoupled cloud region around $r$ $\sim$ 4.5 kpc represent different evolutionary stages of the ram pressure interaction. The initial stage is gas compression caused by ram pressure acceleration of the gas, leading to triggered star formation. Feedback from star formation then facilitates the stripping of most of the remaining molecular gas, probably by disrupting the GMCs and driving the molecular gas to a more diffuse state with lower surface density that is easier to strip. We find evidence for this evolutionary progression from the distribution of young stars in the plume region. Within the bottom part of the plume, there is a ridge seen in H$\alpha$ that angles upward from the disk and is offset from the major axis in the direction of ram pressure. This suggests that the gas in the plume was previously compressed and underwent star formation, and that the kinematic disturbance that caused the compression also gave the star-forming clouds an upward vertical velocity. Consequently, this vertical acceleration imparted to the star-forming clouds by ram pressure compression should make a young thick disk stellar component. Given more time, the gas in the compressed zone should follow a similar path, moving in the same direction from the disk plane as the gas in the H$\alpha$ ridge in the plume region likely did.
	
    \item[$\bullet$ How so much molecular gas is stripped] Estimates for the ram pressure experienced by NGC 4402 suggest it is too weak to directly strip GMCs with surface densities as low as $\Sigma_{gas} \approx 30$ M$_{\odot}$ pc$^{-2}$, yet we find very little dense gas remaining in the stripped region south of the plume. In light of this, we have considered different possible explanations for the large amount of stripped molecular gas in the plume region. Our favored explanation is that GMC lifetimes are shorter than the ram pressure timescale ($\sim$60 Myr) in NGC 4402, thus gas cycles rapidly cycles between GMC and non-GMC phases, and is stripped when it is in a lower density non-GMC state. GMCs may be disrupted by massive star formation on a time scale of around 10 Myr, after which the former GMC density gas ejected by stellar feedback could be stripped away.
	
	\item[$\bullet$ The interrelationship of RPS and SF] Our study illuminates the interrelationships between ram pressure and star formation. Ram pressure can compress the ISM, leading to a locally enhanced rate of star formation. Then feedback from this star formation can make the gas easier to strip, by reducing the gas surface density.
	
\end{description}

\acknowledgments

We gratefully acknowledge support from the NRAO Student Observing Award Program, under award number SOSPA4-003, and from the Joint ALMA Observatory Visitor Program. This paper makes use of the following ALMA data: ADS/JAO.ALMA\#2015.1.01056.S \& 2016.1.00912.S. ALMA is a partnership of ESO (representing its member states), NSF (USA) and NINS (Japan), together with NRC (Canada), MOST and ASIAA (Taiwan), and KASI (Republic of Korea), in cooperation with the Republic of Chile. The Joint ALMA Observatory is operated by ESO, AUI/NRAO and NAOJ. The National Radio Astronomy Observatory is a facility of the National Science Foundation operated under cooperative agreement by Associated Universities, Inc. Based on observations made with the NASA/ESA Hubble Space Telescope, and obtained from the Hubble Legacy Archive, which is a collaboration between the Space Telescope Science Institute (STScI/NASA), the Space Telescope European Coordinating Facility (ST-ECF/ESA) and the Canadian Astronomy Data Centre (CADC/NRC/CSA). We thank the anonymous referee for comments which helped us improve the manuscript.

\bibliographystyle{apalike}
\bibliography{Bibliography}

\begin{thebibliography}{}

\bibitem[{Abramson} et~al., 2016]{Abramson+16}
{Abramson}, A., {Kenney}, J., {Crowl}, H., and {Tal}, T. (2016).
\newblock {HST Imaging of Dust Structures and Stars in the Ram Pressure
  Stripped Virgo Spirals NGC 4402 and NGC 4522: Stripped from the Outside In
  with Dense Cloud Decoupling}.
\newblock {\em \aj}, 152:32.

\bibitem[{Abramson} and {Kenney}, 2014]{Abramson+14}
{Abramson}, A. and {Kenney}, J.~D.~P. (2014).
\newblock {Hubble Space Telescope Imaging of Decoupled Dust Clouds in the Ram
  Pressure Stripped Virgo Spirals NGC 4402 and NGC 4522}.
\newblock {\em \aj}, 147:63.

\bibitem[{Abramson} et~al., 2011]{Abramson+11}
{Abramson}, A., {Kenney}, J.~D.~P., {Crowl}, H.~H., {Chung}, A., {van Gorkom},
  J.~H., {Vollmer}, B., and {Schiminovich}, D. (2011).
\newblock {Caught in the Act: Strong, Active Ram Pressure Stripping in Virgo
  Cluster Spiral NGC 4330}.
\newblock {\em \aj}, 141:164.

\bibitem[{Binggeli} et~al., 1993]{Binggeli+93}
{Binggeli}, B., {Popescu}, C.~C., and {Tammann}, G.~A. (1993).
\newblock {The kinematics of the Virgo cluster revisited}.
\newblock {\em \aaps}, 98:275--296.

\bibitem[{Blitz} et~al., 2007]{Blitz+06}
{Blitz}, L., {Fukui}, Y., {Kawamura}, A., {Leroy}, A., {Mizuno}, N., and
  {Rosolowsky}, E. (2007).
\newblock {Giant Molecular Clouds in Local Group Galaxies}.
\newblock In {Reipurth}, B., {Jewitt}, D., and {Keil}, K., editors, {\em
  Protostars and Planets V}, page~81.

\bibitem[{B{\"o}hringer} et~al., 1994]{Bohringer+94}
{B{\"o}hringer}, H., {Briel}, U.~G., {Schwarz}, R.~A., {Voges}, W., {Hartner},
  G., and {Tr{\"u}mper}, J. (1994).
\newblock {The structure of the Virgo cluster of galaxies from Rosat X-ray
  images}.
\newblock {\em \nat}, 368(6474):828--831.

\bibitem[{Bolatto} et~al., 2013]{Bolatto+13}
{Bolatto}, A.~D., {Wolfire}, M., and {Leroy}, A.~K. (2013).
\newblock {The CO-to-H$_{2}$ Conversion Factor}.
\newblock {\em \araa}, 51(1):207--268.

\bibitem[{Boselli} et~al., 2019]{Boselli+19}
{Boselli}, {Epinat}, {Contini}, {Abril-Melgarejo}, {Boogaard}, {Pointecouteau},
  {Ventou}, {Brinchmann}, {Carton}, {Finley}, {Michel-Dansac}, {Soucail}, and
  {Weilbacher} (2019).
\newblock {Evidence for ram pressure stripping in a cluster of galaxies at
  z=0.7}.
\newblock {\em arXiv e-prints}, page arXiv:1909.05491.

\bibitem[{Boselli} et~al., 2014]{Boselli+14}
{Boselli}, A., {Cortese}, L., {Boquien}, M., {Boissier}, S., {Catinella}, B.,
  {Gavazzi}, G., {Lagos}, C., and {Saintonge}, A. (2014).
\newblock {Cold gas properties of the Herschel Reference Survey. III. Molecular
  gas stripping in cluster galaxies}.
\newblock {\em \aap}, 564:A67.

\bibitem[{Cayatte} et~al., 1990]{Cayatte+90}
{Cayatte}, V., {van Gorkom}, J.~H., {Balkowski}, C., and {Kotanyi}, C. (1990).
\newblock {VLA observations of neutral hydrogen in Virgo Cluster galaxies. I -
  The Atlas}.
\newblock {\em \aj}, 100:604--634.

\bibitem[{Chung} et~al., 2009]{Chung+09}
{Chung}, A., {van Gorkom}, J.~H., {Kenney}, J.~D.~P., {Crowl}, H., and
  {Vollmer}, B. (2009).
\newblock {VLA Imaging of Virgo Spirals in Atomic Gas (VIVA). I. The Atlas and
  the H I Properties}.
\newblock {\em \aj}, 138:1741--1816.

\bibitem[{Cortese} et~al., 2016]{Cortese+16}
{Cortese}, L., {Bekki}, K., {Boselli}, A., {Catinella}, B., {Ciesla}, L.,
  {Hughes}, T.~M., {Baes}, M., {Bendo}, G.~J., {Boquien}, M., {de Looze}, I.,
  {Smith}, M.~W.~L., {Spinoglio}, L., and {Viaene}, S. (2016).
\newblock {The selective effect of environment on the atomic and molecular
  gas-to-dust ratio of nearby galaxies in the Herschel Reference Survey}.
\newblock {\em \mnras}, 459(4):3574--3584.

\bibitem[{Cramer} et~al., 2019]{Cramer+19}
{Cramer}, W.~J., {Kenney}, J.~D.~P., {Sun}, M., {Crowl}, H., {Yagi}, M.,
  {J{\'a}chym}, P., {Roediger}, E., and {Waldron}, W. (2019).
\newblock {Spectacular Hubble Space Telescope Observations of the Coma Galaxy
  D100 and Star Formation in Its Ram Pressure Stripped Tail}.
\newblock {\em \apj}, 870:63.

\bibitem[{Crowl} and {Kenney}, 2006]{Crowl+06}
{Crowl}, H.~H. and {Kenney}, J. D.~P. (2006).
\newblock {The Stellar Population of Stripped Cluster Spiral NGC 4522: A Local
  Analog to K+A Galaxies?}
\newblock {\em \apjl}, 649(2):L75--L78.

\bibitem[{Crowl} et~al., 2005]{Crowl+05}
{Crowl}, H.~H., {Kenney}, J.~D.~P., {van Gorkom}, J.~H., and {Vollmer}, B.
  (2005).
\newblock {Extra-planar Gas and Dust due to Ram Pressure Stripping of the Virgo
  Spiral NGC 4402}.
\newblock In {Braun}, R., editor, {\em Extra-Planar Gas}, volume 331 of {\em
  Astronomical Society of the Pacific Conference Series}, page 281.

\bibitem[{Fossati} et~al., 2018]{Fossati+18}
{Fossati}, M., {Mendel}, J.~T., {Boselli}, A., {Cuillandre}, J.~C., {Vollmer},
  B., {Boissier}, S., {Consolandi}, G., {Ferrarese}, L., {Gwyn}, S., {Amram},
  P., {Boquien}, M., {Buat}, V., {Burgarella}, D., {Cortese}, L.,
  {C{\^o}t{\'e}}, P., {C{\^o}t{\'e}}, S., {Durrell}, P., {Fumagalli}, M.,
  {Gavazzi}, G., {Gomez-Lopez}, J., {Hensler}, G., {Koribalski}, B.,
  {Longobardi}, A., {Peng}, E.~W., {Roediger}, J., {Sun}, M., and {Toloba}, E.
  (2018).
\newblock {A Virgo Environmental Survey Tracing Ionised Gas Emission (VESTIGE).
  II. Constraining the quenching time in the stripped galaxy NGC 4330}.
\newblock {\em \aap}, 614:A57.

\bibitem[{Gil de Paz} et~al., 2007]{Paz+07}
{Gil de Paz}, A., {Boissier}, S., {Madore}, B.~F., {Seibert}, M., {Joe}, Y.~H.,
  {Boselli}, A., {Wyder}, T.~K., {Thilker}, D., {Bianchi}, L., {Rey}, S.-C.,
  {Rich}, R.~M., {Barlow}, T.~A., {Conrow}, T., {Forster}, K., {Friedman},
  P.~G., {Martin}, D.~C., {Morrissey}, P., {Neff}, S.~G., {Schiminovich}, D.,
  {Small}, T., {Donas}, J., {Heckman}, T.~M., {Lee}, Y.-W., {Milliard}, B.,
  {Szalay}, A.~S., and {Yi}, S. (2007).
\newblock {The GALEX Ultraviolet Atlas of Nearby Galaxies}.
\newblock {\em \apjs}, 173(2):185--255.

\bibitem[{Giovanelli} and {Haynes}, 1983]{Giovanelli+83}
{Giovanelli}, R. and {Haynes}, M.~P. (1983).
\newblock {The H I extent and deficiency of spiral galaxies in the Virgo
  cluster}.
\newblock {\em \aj}, 88:881--908.

\bibitem[{Gunn} and {Gott}, 1972]{Gunn+72}
{Gunn}, J.~E. and {Gott}, III, J.~R. (1972).
\newblock {On the Infall of Matter Into Clusters of Galaxies and Some Effects
  on Their Evolution}.
\newblock {\em \apj}, 176:1.

\bibitem[{Heyer} and {Dame}, 2015]{Heyer+15}
{Heyer}, M. and {Dame}, T.~M. (2015).
\newblock {Molecular Clouds in the Milky Way}.
\newblock {\em \araa}, 53:583--629.

\bibitem[{J{\'a}chym} et~al., 2014]{Jachym+14}
{J{\'a}chym}, P., {Combes}, F., {Cortese}, L., {Sun}, M., and {Kenney},
  J.~D.~P. (2014).
\newblock {Abundant Molecular Gas and Inefficient Star Formation in
  Intracluster Regions: Ram Pressure Stripped Tail of the Norma Galaxy
  ESO137-001}.
\newblock {\em \apj}, 792:11.

\bibitem[{J{\'a}chym} et~al., 2017]{Jachym+17}
{J{\'a}chym}, P., {Sun}, M., {Kenney}, J.~D.~P., {Cortese}, L., {Combes}, F.,
  {Yagi}, M., {Yoshida}, M., {Palou{\v s}}, J., and {Roediger}, E. (2017).
\newblock {Molecular Gas Dominated 50 kpc Ram Pressure Stripped Tail of the
  Coma Galaxy D100}.
\newblock {\em \apj}, 839:114.

\bibitem[{J{\'o}zsa} et~al., 2007]{Jozsa+07}
{J{\'o}zsa}, G.~I.~G., {Kenn}, F., {Klein}, U., and {Oosterloo}, T.~A. (2007).
\newblock {Kinematic modelling of disk galaxies. I. A new method to fit tilted
  rings to data cubes}.
\newblock {\em \aap}, 468:731--774.

\bibitem[{Kenney} et~al., 2015]{Kenney+15}
{Kenney}, J.~D.~P., {Abramson}, A., and {Bravo-Alfaro}, H. (2015).
\newblock {Hubble Space Telescope and HI Imaging of Strong Ram Pressure
  Stripping in the Coma Spiral NGC 4921: Dense Cloud Decoupling and Evidence
  for Magnetic Binding in the ISM}.
\newblock {\em \aj}, 150:59.

\bibitem[{Kenney} et~al., 2014]{Kenney+14}
{Kenney}, J.~D.~P., {Geha}, M., {J{\'a}chym}, P., {Crowl}, H.~H., {Dague}, W.,
  {Chung}, A., {van Gorkom}, J., and {Vollmer}, B. (2014).
\newblock {Transformation of a Virgo Cluster Dwarf Irregular Galaxy by Ram
  Pressure Stripping: IC3418 and Its Fireballs}.
\newblock {\em \apj}, 780:119.

\bibitem[{Kenney} et~al., 2008]{Kenney+08}
{Kenney}, J. D.~P., {Tal}, T., {Crowl}, H.~H., {Feldmeier}, J., and {Jacoby},
  G.~H. (2008).
\newblock {A Spectacular H{\ensuremath{\alpha}} Complex in Virgo: Evidence for
  a Collision between M86 and NGC 4438 and Implications for the Collisional ISM
  Heating of Ellipticals}.
\newblock {\em \apj}, 687(2):L69.

\bibitem[{Kenney} et~al., 2004]{Kenney+04}
{Kenney}, J. D.~P., {van Gorkom}, J.~H., and {Vollmer}, B. (2004).
\newblock {VLA H I Observations of Gas Stripping in the Virgo Cluster Spiral
  NGC 4522}.
\newblock {\em \aj}, 127(6):3361--3374.

\bibitem[{Kenney} et~al., 1992]{Kenney+92}
{Kenney}, J.~D.~P., {Wilson}, C.~D., {Scoville}, N.~Z., {Devereux}, N.~A., and
  {Young}, J.~S. (1992).
\newblock {Twin peaks of CO emission in the central regions of barred
  galaxies}.
\newblock {\em \apjl}, 395:L79--L82.

\bibitem[{Kenney} et~al., 2009]{Kenney+09}
{Kenney}, J.~D.~P., {Wong}, O.~I., {Abramson}, A., {Howell}, J.~H., {Murphy},
  E.~J., and {Helou}, G.~X. (2009).
\newblock {Spitzer Observations of Environmental Effects on Virgo Cluster
  Galaxies}.
\newblock In {\em The Evolving ISM in the Milky Way and Nearby Galaxies},
  page~9.

\bibitem[{Kenney} and {Young}, 1989]{Kenney+89}
{Kenney}, J.~D.~P. and {Young}, J.~S. (1989).
\newblock {The effects of environment on the molecular and atomic gas
  properties of large Virgo cluster spirals}.
\newblock {\em \apj}, 344:171--199.

\bibitem[{Klessen} and {Glover}, 2016]{Klessen+16}
{Klessen}, R.~S. and {Glover}, S. C.~O. (2016).
\newblock {Physical Processes in the Interstellar Medium}.
\newblock {\em Saas-Fee Advanced Course}, 43:85.

\bibitem[{Koopmann} and {Kenney}, 2004]{Koopmann+04}
{Koopmann}, R.~A. and {Kenney}, J.~D.~P. (2004).
\newblock {H{$\alpha$} Morphologies and Environmental Effects in Virgo Cluster
  Spiral Galaxies}.
\newblock {\em \apj}, 613:866--885.

\bibitem[{Kruijssen} et~al., 2019]{Kruijssen+19}
{Kruijssen}, J.~M.~D., {Schruba}, A., {Chevance}, M., {Longmore}, S.~N.,
  {Hygate}, A. e. P.~S., {Haydon}, D.~T., {McLeod}, A.~F., {Dalcanton}, J.~J.,
  {Tacconi}, L.~J., and {van Dishoeck}, E.~F. (2019).
\newblock {Fast and inefficient star formation due to short-lived molecular
  clouds and rapid feedback}.
\newblock {\em \nat}, 569(7757):519--522.

\bibitem[{Lee} et~al., 2017]{Lee+17}
{Lee}, B., {Chung}, A., {Tonnesen}, S., {Kenney}, J.~D.~P., {Wong}, O.~I.,
  {Vollmer}, B., {Petitpas}, G.~R., {Crowl}, H.~H., and {van Gorkom}, J.
  (2017).
\newblock {The effect of ram pressure on the molecular gas of galaxies: three
  case studies in the Virgo cluster}.
\newblock {\em \mnras}, 466:1382--1398.

\bibitem[{Lee} and {Jang}, 2016]{Lee+16}
{Lee}, M.~G. and {Jang}, I.~S. (2016).
\newblock {Globular Clusters and Spur Clusters in NGC 4921, the Brightest
  Spiral Galaxy in the Coma Cluster}.
\newblock {\em \apj}, 819:77.

\bibitem[{Leroy} et~al., 2009]{Leroy+09}
{Leroy}, A.~K., {Walter}, F., {Bigiel}, F., {Usero}, A., {Weiss}, A., {Brinks},
  E., {de Blok}, W.~J.~G., {Kennicutt}, R.~C., {Schuster}, K.-F., {Kramer}, C.,
  {Wiesemeyer}, H.~W., and {Roussel}, H. (2009).
\newblock {Heracles: The HERA CO Line Extragalactic Survey}.
\newblock {\em \aj}, 137:4670--4696.

\bibitem[{Mayer} et~al., 2006]{Mayer+06}
{Mayer}, L., {Mastropietro}, C., {Wadsley}, J., {Stadel}, J., and {Moore}, B.
  (2006).
\newblock {Simultaneous ram pressure and tidal stripping; how dwarf spheroidals
  lost their gas}.
\newblock {\em \mnras}, 369(3):1021--1038.

\bibitem[{Mei} et~al., 2007]{Mei+07}
{Mei}, S., {Blakeslee}, J.~P., {C{\^o}t{\'e}}, P., {Tonry}, J.~L., {West},
  M.~J., {Ferrarese}, L., {Jord{\'a}n}, A., {Peng}, E.~W., {Anthony}, A., and
  {Merritt}, D. (2007).
\newblock {The ACS Virgo Cluster Survey. XIII. SBF Distance Catalog and the
  Three-dimensional Structure of the Virgo Cluster}.
\newblock {\em \apj}, 655(1):144--162.

\bibitem[{Merluzzi} et~al., 2016]{Merluzzi+16}
{Merluzzi}, P., {Busarello}, G., {Dopita}, M.~A., {Haines}, C.~P.,
  {Steinhauser}, D., {Bourdin}, H., and {Mazzotta}, P. (2016).
\newblock {Shapley Supercluster Survey: ram-pressure stripping versus tidal
  interactions in the Shapley supercluster}.
\newblock {\em \mnras}, 460:3345--3369.

\bibitem[{Moon} et~al., 2019]{Moon+19}
{Moon}, J.-S., {An}, S.-H., and {Yoon}, S.-J. (2019).
\newblock {Living with Neighbors. I. Observational Clues to Hydrodynamic Impact
  of Neighboring Galaxies on Star Formation}.
\newblock {\em \apj}, 882:14.

\bibitem[{Moretti} et~al., 2020]{Moretti+20}
{Moretti}, A., {Paladino}, R., {Poggianti}, B.~M., {Serra}, P., {Roediger}, E.,
  {Gullieuszik}, M., {Tomi{\v{c}}i{\'c}}, N., {Radovich}, M., {Vulcani}, B.,
  {Jaff{\'e}}, Y.~L., {Fritz}, J., {Bettoni}, D., {Ramatsoku}, M., and
  {Wolter}, A. (2020).
\newblock {GASP. XXII. The Molecular Gas Content of the JW100 Jellyfish Galaxy
  at z {\ensuremath{\sim}} 0.05: Does Ram Pressure Promote Molecular Gas
  Formation?}
\newblock {\em \apj}, 889(1):9.

\bibitem[{Murphy} et~al., 2009]{Murphy+09}
{Murphy}, E.~J., {Kenney}, J.~D.~P., {Helou}, G., {Chung}, A., and {Howell},
  J.~H. (2009).
\newblock {Environmental Effects in Clusters: Modified Far-Infrared-Radio
  Relations within Virgo Cluster Galaxies}.
\newblock {\em \apj}, 694:1435--1451.

\bibitem[{Pappalardo} et~al., 2010]{Pappalardo+10}
{Pappalardo}, C., {Lan{\c c}on}, A., {Vollmer}, B., {Ocvirk}, P., {Boissier},
  S., and {Boselli}, A. (2010).
\newblock {Pinning down the ram-pressure-induced halt of star formation in the
  Virgo cluster spiral galaxy NGC 4388. A joint inversion of spectroscopic and
  photometric data}.
\newblock {\em \aap}, 514:A33.

\bibitem[{Pety} et~al., 2013]{Pety+13}
{Pety}, J., {Schinnerer}, E., {Leroy}, A.~K., {Hughes}, A., {Meidt}, S.~E.,
  {Colombo}, D., {Dumas}, G., {Garc{\'\i}a-Burillo}, S., {Schuster}, K.~F., and
  {Kramer}, C. (2013).
\newblock {The Plateau de Bure + 30 m Arcsecond Whirlpool Survey Reveals a
  Thick Disk of Diffuse Molecular Gas in the M51 Galaxy}.
\newblock {\em \apj}, 779(1):43.

\bibitem[{Poggianti} et~al., 2019]{Poggianti+19c}
{Poggianti}, B.~M., {Gullieuszik}, M., {Tonnesen}, S., {Moretti}, A.,
  {Vulcani}, B., {Radovich}, M., {Jaff{\'e}}, Y., {Fritz}, J., {Bettoni}, D.,
  {Franchetto}, A., {Fasano}, G., {Bellhouse}, C., and {Omizzolo}, A. (2019).
\newblock {GASP XIII. Star formation in gas outside galaxies}.
\newblock {\em \mnras}, 482(4):4466--4502.

\bibitem[{Quilis} et~al., 2000]{Quilis+00}
{Quilis}, V., {Moore}, B., and {Bower}, R. (2000).
\newblock {Gone with the Wind: The Origin of S0 Galaxies in Clusters}.
\newblock {\em Science}, 288:1617--1620.

\bibitem[{Roberts} and {Parker}, 2020]{Roberts+20}
{Roberts}, I.~D. and {Parker}, L.~C. (2020).
\newblock {Ram pressure stripping candidates in the Coma Cluster: Evidence for
  enhanced star formation}.
\newblock {\em arXiv e-prints}, page arXiv:2004.12033.

\bibitem[{Roman-Duval} et~al., 2016]{Roman+16}
{Roman-Duval}, J., {Heyer}, M., {Brunt}, C.~M., {Clark}, P., {Klessen}, R., and
  {Shetty}, R. (2016).
\newblock {Distribution and Mass of Diffuse and Dense CO Gas in the Milky Way}.
\newblock {\em \apj}, 818(2):144.

\bibitem[{Schindler} et~al., 1999]{Schindler+99}
{Schindler}, S., {Binggeli}, B., and {B{\"o}hringer}, H. (1999).
\newblock {Morphology of the Virgo cluster: Gas versus galaxies}.
\newblock {\em \aap}, 343:420--438.

\bibitem[{Sellwood} and {Spekkens}, 2015]{Sellwood+15}
{Sellwood}, J.~A. and {Spekkens}, K. (2015).
\newblock {DiskFit: a code to fit simple non-axisymmetric galaxy models either
  to photometric images or to kinematic maps}.
\newblock {\em arXiv e-prints}.

\bibitem[{Smethurst} et~al., 2017]{Smethurst+17}
{Smethurst}, R.~J., {Lintott}, C.~J., {Bamford}, S.~P., {Hart}, R.~E., {Kruk},
  S.~J., {Masters}, K.~L., {Nichol}, R.~C., and {Simmons}, B.~D. (2017).
\newblock {Galaxy Zoo: the interplay of quenching mechanisms in the group
  environment}.
\newblock {\em \mnras}, 469(3):3670--3687.

\bibitem[{Solomon} et~al., 1997]{Solomon+97}
{Solomon}, P.~M., {Downes}, D., {Radford}, S.~J.~E., and {Barrett}, J.~W.
  (1997).
\newblock {The Molecular Interstellar Medium in Ultraluminous Infrared
  Galaxies}.
\newblock {\em \apj}, 478(1):144--161.

\bibitem[{Sun} et~al., 2010]{Sun+10}
{Sun}, M., {Donahue}, M., {Roediger}, E., {Nulsen}, P.~E.~J., {Voit}, G.~M.,
  {Sarazin}, C., {Forman}, W., and {Jones}, C. (2010).
\newblock {Spectacular X-ray Tails, Intracluster Star Formation, and ULXs in
  A3627}.
\newblock {\em \apj}, 708:946--964.

\bibitem[{Tonnesen} and {Bryan}, 2009]{Tonnesen+09}
{Tonnesen}, S. and {Bryan}, G.~L. (2009).
\newblock {Gas Stripping in Simulated Galaxies with a Multiphase Interstellar
  Medium}.
\newblock {\em \apj}, 694:789--804.

\bibitem[{Tonnesen} and {Bryan}, 2010]{Tonnesen+10}
{Tonnesen}, S. and {Bryan}, G.~L. (2010).
\newblock {The Tail of the Stripped Gas that Cooled: H I, H{$\alpha$}, and
  X-ray Observational Signatures of Ram Pressure Stripping}.
\newblock {\em \apj}, 709:1203--1218.

\bibitem[{Tonnesen} and {Bryan}, 2012]{Tonnesen+12}
{Tonnesen}, S. and {Bryan}, G.~L. (2012).
\newblock {Star formation in ram pressure stripped galactic tails}.
\newblock {\em \mnras}, 422:1609--1624.

\bibitem[{Vogel} et~al., 1988]{Vogel+88}
{Vogel}, S.~N., {Kulkarni}, S.~R., and {Scoville}, N.~Z. (1988).
\newblock {Star formation in giant molecular associations synchronized by a
  spiral density wave}.
\newblock {\em \nat}, 334(6181):402--406.

\bibitem[{Vollmer} et~al., 2005]{Vollmer+05}
{Vollmer}, B., {Braine}, J., {Combes}, F., and {Sofue}, Y. (2005).
\newblock {New CO observations and simulations of the NGC 4438/NGC 4435 system.
  Interaction diagnostics of the Virgo cluster galaxy NGC 4438}.
\newblock {\em \aap}, 441(2):473--489.

\bibitem[{Vollmer} et~al., 2008]{Vollmer+08}
{Vollmer}, B., {Braine}, J., {Pappalardo}, C., and {Hily-Blant}, P. (2008).
\newblock {Ram-pressure stripped molecular gas in the Virgo spiral galaxy NGC
  4522}.
\newblock {\em \aap}, 491:455--464.

\bibitem[{Vollmer} et~al., 2001]{Vollmer+01a}
{Vollmer}, B., {Cayatte}, V., {Balkowski}, C., and {Duschl}, W.~J. (2001).
\newblock {Ram Pressure Stripping and Galaxy Orbits: The Case of the Virgo
  Cluster}.
\newblock {\em \apj}, 561:708--726.

\bibitem[{Vollmer} et~al., 2012a]{Vollmer+12d}
{Vollmer}, B., {Soida}, M., {Braine}, J., {Abramson}, A., {Beck}, R., {Chung},
  A., {Crowl}, H.~H., {Kenney}, J.~D.~P., and {van Gorkom}, J.~H. (2012a).
\newblock {Ram pressure stripping of the multiphase ISM and star formation in
  the Virgo spiral galaxy NGC 4330}.
\newblock {\em \aap}, 537:A143.

\bibitem[{Vollmer} et~al., 2012b]{Vollmer+12}
{Vollmer}, B., {Wong}, O.~I., {Braine}, J., {Chung}, A., and {Kenney}, J.~D.~P.
  (2012b).
\newblock {The influence of the cluster environment on the star formation
  efficiency of 12 Virgo spiral galaxies}.
\newblock {\em \aap}, 543:A33.

\bibitem[{Vulcani} et~al., 2018]{Vulcani+18}
{Vulcani}, B., {Poggianti}, B.~M., {Gullieuszik}, M., {Moretti}, A.,
  {Tonnesen}, S., {Jaff{\'e}}, Y.~L., {Fritz}, J., {Fasano}, G., and {Bettoni},
  D. (2018).
\newblock {Enhanced Star Formation in Both Disks and Ram-pressure-stripped
  Tails of GASP Jellyfish Galaxies}.
\newblock {\em \apj}, 866:L25.

\end{thebibliography}

\end{document}